\documentclass[a4paper,11pt]{article}
\usepackage[utf8]{inputenc}
\usepackage[english]{babel}
\usepackage{braket}
\usepackage{xspace}
\usepackage{bbm}
\usepackage{mathdots}
\usepackage{stackrel}
\usepackage{xcolor}
\usepackage{mathtools}
\usepackage{bbm}
\usepackage{subfigure}
\usepackage[T1]{fontenc}               %
\usepackage[left=3cm,
            right=2.5cm,
            top=2.5cm,
            bottom=3cm
            ]{geometry}                %
\usepackage{graphicx}
\usepackage{mathrsfs}
\usepackage{appendix}
\usepackage{fancyhdr}
\newcommand{\ee}{\mathrm{e}}
\usepackage{amsmath,                   %
            amssymb,                   %
            amsthm}                    %
\usepackage{cite}
\usepackage{setspace} 
\usepackage[colorlinks=true,linkcolor=blue, citecolor=blue, bookmarks]{hyperref}
\usepackage[protrusion=true,expansion=true]{microtype}

\usepackage{authblk}

\theoremstyle{definition}

\theoremstyle{remark}

\def\RR{\mathbb{R}}

\def\ii{\mathrm{i}}

\DeclareMathOperator{\Tr}{Tr}
\DeclareMathOperator{\Id}{Id}

\date{}\title{Entanglement and negativity Hamiltonians \\ for the massless Dirac field on the half line}

\author[1]{Federico Rottoli}
\author[1,2]{Sara Murciano}
\author[1]{Erik Tonni}
\author[1,3]{Pasquale Calabrese}

\affil[1]{SISSA and INFN Sezione di Trieste, via Bonomea 265, 34136 Trieste, Italy.}
\affil[2]{Walter Burke Institute for Theoretical Physics, California
Institute of Technology, Pasadena, California 91125, USA.}
\affil[3]{International Centre for Theoretical Physics (ICTP), Strada Costiera 11, 34151 Trieste, Italy.}

\begin{document}
\maketitle

\begin{abstract}
We study the ground-state entanglement Hamiltonian of several disjoint intervals for the massless Dirac fermion on the half-line.
Its structure consists of a local part and a bi-local term that couples each point to another one in each other interval. 
The bi-local operator can be either diagonal or mixed in the fermionic chiralities and it is sensitive to the boundary conditions.
The knowledge of such entanglement Hamiltonian is the starting point to evaluate the negativity Hamiltonian, i.e. the logarithm of the 
partially transposed reduced density matrix, which is an operatorial characterisation of entanglement of subsystems in a mixed states.
We find that the negativity Hamiltonian inherits the structure of the corresponding entanglement Hamiltonian.
We finally show how the continuum expressions for both these operators can be recovered from 
exact numerical computations in free-fermion chains.
\end{abstract}

\newpage

\tableofcontents
 
\section{Introduction}

As Schr\"odinger recognised, entanglement is \emph{the} characteristic feature of quantum mechanics.
In recent years, the study of quantum entanglement in many-body systems has been
so fruitful that it has led to an unprecedented exchange of ideas and concepts among previously unrelated fields,
such as quantum information \cite{nc-10}, high-energy physics \cite{nrt-09,Rangamani}, statistical mechanics \cite{intro1,intro2,eisert-2010}, condensed matter \cite{intro3} and many more. 
In these different contexts, the most popular and successful measure of entanglement for bipartite pure systems $A \cup B$ are the R\'enyi entropies. Starting from a pure state $\ket{\psi}$
belonging to the factorised Hilbert space $\mathcal{H}=\mathcal{H}_A \otimes \mathcal{H}_B$, one  
defines the reduced density matrix (RDM) as $\hat{\rho}_A=\mathrm{Tr}_B \ket{\psi}\bra{\psi}$. This is the key object to define the R\'enyi entropies
\begin{equation}\label{eq:renyi}
    S^{(m)}_A=\frac{1}{1-m}\log\Tr_A\hat{\rho}_A^m,
\end{equation}
where $\textrm{Tr}_A$ will be denoted simply by $\textrm{Tr}$ hereafter.
Despite being powerful measures of quantum correlations, both from a theoretical and experimental point of view (see e.g. \cite{cc-04,cw-94,hlw-94,kaufman-2016,brydges-2019,fis}), the R\'enyi entropies sometimes cannot provide a complete characterisation of entanglement. A more satisfactory understanding is provided by operatorial quantities.
Being $\hat \rho_A$ hermitian, positive semi-definite and normalised to $\Tr \hat \rho_A=1$, it can be written as
\begin{equation}\label{eq:KAintro}
    \hat{\rho}_A=\frac{e^{-2\pi \hat{K}_A}}{Z_A}, \qquad Z_A=\Tr e^{-2\pi \hat{K}_A},
\end{equation}
where the hermitian operator $\hat{K}_A$ is called \emph{entanglement} (or \emph{modular}) \emph{Hamiltonian}. Both the reduced density matrix and
the entanglement Hamiltonian provide the complete information about the bipartite entanglement between a subsystem $A$ and its complement. However, contrarily to the R\'enyi entropies, which are solely sensitive to the eigenspectrum of the RDM (the entanglement spectrum \cite{Li2008}), the finer structure of the entanglement Hamiltonian, which strictly depends on the eigenvectors of the RDM and their relation to the corresponding eigenvalues, is considerably more challenging to characterise at the theoretical
and experimental level.
Nevertheless, we can identify at least three research streams which stimulated the study of this operator by different communities. 
The first result arose in algebraic quantum field theory during the 70s: it is a theorem by Bisognano and Wichmann, which proved 
the local nature of the entanglement Hamiltonian for the ground state of a relativistic quantum field theory (QFT) in $(D+1)$ dimensions
when the subsystem is the half-space $A = \left \{x \in \RR^{D+1} \big{|} x^1 > 0,  x^0 = t = 0 \right \}$.
In this case the entanglement Hamiltonian is the generator of the Lorentz transformations \cite{bw-75,bw-76}
\begin{equation}\label{eq:BWth}
    \hat{K}_A = \int_A d^D\! x\, x^1\, T_{00} (x),
\end{equation}
where $T_{00}$ is the energy density. In turns, this result provides a physical mechanism for the origin of the celebrated Unruh effect \cite{Unruh1976}. 
While the Bisognano-Wichmann theorem applies to Lorentz invariant QFTs in any dimension for the half-space, in $1+1$ the conformal symmetry of a system allows to map
\eqref{eq:BWth} to more general geometries. For the simplest bipartition where $A$ is a single interval of length $\ell$ on an infinite line and the conformal field theory (CFT) is in its ground state, 
the entanglement Hamiltonian reads \cite{HislopLongo82, Casini2011, KlichWong2013, ct}
\begin{equation}\label{eq:oneInterval}
    \hat{K}_A=\int_A dx\, \frac{x \left (\ell - x \right )}{\ell}\, T_{00}(x) .
\end{equation}

During the course of the 90s, a general construction $\hat{K}_A$ for Gaussian states arising in lattice models of statistical mechanics was provided \cite{pkl-99,Gaussian}, 
which was the starting point for a plethora of subsequent computations of the entanglement Hamiltonians in these lattice systems 
\cite{Peschel2009, Chung2001, Peschel2003, Peschel2004, Peschel2012, Giudici2018,zdc-2020,EH-l2,EH-l3}.
A final and fundamental use of the entanglement Hamiltonian is the identification of topological phases of matter thanks to the Li-Haldane conjecture \cite{Li2008}, which 
turned out to be a much more effective tool than the R\'enyi entropies \cite{lw-05, kp-06}.

More recently, an operator playing the same role as the entanglement Hamiltonian but for mixed states (or equivalently for tripartitions of pure states) has been introduced in \cite{mvdc-22}. Also in this context, one can measure the entanglement using a scalar quantity defined from the partial transpose of the RDM. 
Considering a further partition of the subsystem as $A = A_1 \cup A_2$, we can write $\rho_A$ as
\begin{equation}
\hat{\rho}_A\, =\sum_{i,j,k,l}\braket{e^1_i,e^2_j|\hat{\rho}_A|e^1_k, e^2_l}\ket{e_i^1,e_j^2}\bra{e^1_k,e^2_l},
\end{equation}
where $\ket{e_j^1}$ and $\ket{e_k^2}$ are orthonormal bases in the Hilbert spaces $\mathcal{H}_1$ and $\mathcal{H}_2$ corresponding to the $A_1$ and $A_2$ regions, 
respectively. The partial transpose of the reduced density matrix is defined by exchanging the matrix elements in, e.g., the subsystem $A_2$
\begin{equation}
\label{eq:bosonic}
\hat{\rho}^{T_2}_A\, =\sum_{i,j,k,l}\braket{e^1_i,e^2_l|\hat{\rho}_A|e^1_k, e^2_j}\ket{e_i^1,e_j^2}\bra{e^1_k,e^2_l}.
\end{equation}
The importance of this operation comes from the Peres-Horodecki criterion \cite{p-96, s-00}, which implies that negative eigenvalues of $\hat{\rho}^{T_2}_A$ signal the presence of entanglement between $A_1$ and $A_2$.
Then, a computable entanglement measure is provided by the logarithmic negativity \cite{vidal,plenio-2005}
\begin{equation}\label{eq:neg_defT2}
    \mathcal{E}\equiv \log \Tr \left |\hat{\rho}_A^{T_2} \right |,
\end{equation}
which has been studied in two dimensional CFT \cite{cct-neg-1, cct-neg-2, cct-neg-3}.
In analogy with \eqref{eq:KAintro}, the negativity Hamiltonian has been defined as \cite{mvdc-22}
\begin{equation}
    \hat{\rho}_A^{T_2}=\frac{e^{-2\pi\hat{N}_A}}{Z_A},
\end{equation}
where now $\hat{N}_A$ is non-hermitian because of the possible negative eigenvalues of $\hat{\rho}_A^{T_2}$.

Up to this point, we have set the stage for an operatorial characterisation of 
entanglement, both in pure and mixed states. The goal of this manuscript is to obtain analytical expressions for the entanglement and negativity Hamiltonians when $A$ consists of an arbitrary number of intervals in the presence of physical boundaries. There are many reasons to investigate these systems: experimental solid-state systems typically have open boundary conditions (OBC);
in trapped cold atoms, the vanishing of the density outside the trap induces OBC in the inhomogeneous gas that can be treated through field theories in curved space \cite{dsvc-17}; in some non-equilibrium protocols like a quantum quench, the initial state can be seen as a boundary state in imaginary time formalism \cite{cc-05}.

The system that we study in this work
is the ground state of a massless Dirac fermion
on the half line $x > 0$, with a boundary condition at $x = 0$ that guarantees 
energy conservation \cite{Cardy84, Cardy86, Cardy89}. 
The boundary condition can be 
implemented by two distinct phases \cite{lm-98, m-11, mt-21, mt2-21}, 
which correspond to models characterised by different conservation laws, 
preserving either charge or helicity but not both of them at the same time. 
As a consequence, this 
induces a coupling between the left and right-moving chiral components of the Dirac field, as we review in Section \ref{sec:reviewDirac}. 
The rest of the manuscript is organised as follows. In Section \ref{sec:EH}, after a review of 
the entanglement Hamiltonian $\hat{K}_A$ for
an arbitrary number of intervals on the full line, we compute $\hat{K}_A$ in the presence of the boundary in 
both phases. In Section \ref{sec:NH}, according to the construction suggested in \cite{mvdc-22}, we then use this result to analyse the negativity Hamiltonian of two intervals on the half-line. Finally,  in Section \ref{sec:lattice}, we
focus on the non-trivial task of obtaining from the entanglement and negativity Hamiltonians of a tight-binding model the corresponding continuum results in the underlying 
Dirac field theory.
We draw our conclusions in Section \ref{sec:conclusions}.

\section{Dirac fermions in the presence of a boundary}\label{sec:reviewDirac}
In this section, we recall the main properties of the massless Dirac fermion 
in the $1+1$ dimensional half Minkowski spacetime corresponding to $x \geqslant 0$,
with particular focus on the effect of different boundary conditions \cite{lm-98, mt-21, mt2-21}.

The complex Dirac fermion is given by the doublet
\begin{equation}\label{eq:doublet}
    \Psi(x,t) = \begin{pmatrix}
        \psi_R(x-t)\\
        \psi_L(x+t)
    \end{pmatrix},
\end{equation}
where the two components are respectively the right- and the left-moving fermions. 
For later convenience, we remind that each of these components can be written in terms of two real Majorana fermions $\mu^{1}$ and $\mu^2$ as 
\begin{equation}\label{eq:DiracMajorana}
    \begin{cases}
        \psi(x, t) = \mu^1(x, t) + \ii \mu^2(x, t)\\
        \psi^\dagger(x, t) = \mu^1(x, t) - \ii \mu^2(x, t) .
    \end{cases}
\end{equation}
Writing the gamma matrices as
\begin{equation}
    \gamma^0 = \begin{pmatrix}
        0&  1\\
        1&  0
    \end{pmatrix},\qquad \gamma^1 = \begin{pmatrix}
        0&  -1\\
        1&  0
    \end{pmatrix},
\end{equation}
the dynamics of the model is described by the Lagrangian density
\begin{equation}\label{eq:DiracLagrangian}\begin{split}
    \mathcal{L}(x,t) =&\, \ii\, \Psi^\dagger(x, t)\, \gamma^0 \gamma^\mu\, \partial_\mu \Psi(x, t) \\
    =&\, \ii\, \psi_R^\dagger(x-t) \left ( \partial_t + \partial_x \right ) \psi_R(x-t) + \ii\, \psi_L^\dagger(x+t) \left ( \partial_t - \partial_x \right ) \psi_L(x+t),
\end{split}\end{equation}
in which, as expected for a CFT, the two chiral components are decoupled. From the Lagrangian \eqref{eq:DiracLagrangian}, we can find the two-point correlation function of the Dirac fermion. For the two chiral fermions, the equal time two-point correlation functions are
\begin{equation}
    \begin{cases}
        \langle \psi_R^\dagger(x-t) \psi_R(y-t) \rangle = C(x-y)\\
        \langle \psi_L^\dagger(x+t) \psi_L(y+t) \rangle= C(-x+y)\,,
    \end{cases}
\end{equation}
where 
\begin{equation}\label{eq:DiracCorr}
    C (x-y) \equiv \frac{1}{2\pi \ii} \frac{1}{(x-y)-\ii\epsilon} = \frac{1}{2}\delta(x-y) - \frac{\ii}{2\pi}\mathcal{P}\frac{1}{x-y},
\end{equation}
and $\mathcal{P}$ denotes the Cauchy principal value. 
We remark that, for the massless Dirac field on the line, 
the two chiralities are decoupled because the two-point correlators involving the components with different chirality vanish.
As we will see in the following section, for free theories the entanglement Hamiltonian can be obtained by the knowledge of the equal time two-point correlation functions.

The Lagrangian \eqref{eq:DiracLagrangian} is invariant under two global transformations: the vector phase transformation
\begin{equation}\label{eq:vectorSymm}
    {\begin{pmatrix}
        \psi_R (x-t)\\
        \psi_L (x+t)
    \end{pmatrix}} \longrightarrow e^{\ii\theta_v} 
    {\begin{pmatrix}
        \psi_R (x-t)\\
        \psi_L (x+t)
    \end{pmatrix}},
\end{equation}
which multiplies both chiralities by the same phase,
and the axial phase transformation
\begin{equation}\label{eq:axialSymm}
    {\begin{pmatrix}
        \psi_R (x-t)\\
        \psi_L (x+t)
    \end{pmatrix}} \longrightarrow \begin{pmatrix}
        e^{-\ii\theta_a}\, \psi_R (x-t)\\
        e^{\ii\theta_a}\, \psi_L (x+t)
    \end{pmatrix},
\end{equation}
which multiplies them by conjugate phases.

When a CFT  is defined on the half line $x \geqslant 0$, 
it is natural to impose boundary conditions that ensure the global energy conservation \cite{Cardy84, Cardy86, Cardy89}.
For the massless Dirac field, such requirement leads to two possible boundary conditions that mix the components with different chirality and 
break either the vector or the axial symmetry \cite{lm-98, mt-21}. Each boundary condition defines a specific model (phase).
We distinguish the two phases according to which bulk symmetry is preserved by the boundary condition.

\paragraph{Vector phase:}
Denoting the massless Dirac field on the half line $x\geqslant 0$ as
\begin{equation}
    \Lambda(x, t) = {\begin{pmatrix}
        \lambda_R(x-t)\\
        \lambda_L(x+t)
    \end{pmatrix}},
\end{equation}
where $\lambda_{R}$ and $\lambda_{L}$ are the two chiral components,
the vector phase is defined by the following family of boundary condition at $x=0$
\begin{equation}\label{eq:vectorCond}
    \lambda_R(t) = e^{\ii\alpha_v} \lambda_L(t),\qquad \alpha_v \in [0, 2\pi)\,.
\end{equation}
In this phase, because of the boundary condition, 
the vector symmetry (\ref{eq:vectorSymm}) is preserved, while the axial symmetry (\ref{eq:axialSymm}) is broken.
The occurrence of a coupling at the boundary between the components of the massless Dirac field
leads to non vanishing correlators for fields having different chirality. 
The equal time correlation matrix in the vector phase reads \cite{mt-21}
\begin{equation}\label{eq:vectorCorr}\begin{split}
    \langle\Lambda(x,t) \Lambda^\dagger(y,t) \rangle
    &= {\begin{pmatrix}
        \langle\lambda_R(x-t) \lambda_R^\dagger(y-t) \rangle &   \langle\lambda_R(x-t) \lambda_L^\dagger(y+t) \rangle \\
              \rule{0pt}{.5cm}
        \langle\lambda_L(x+t) \lambda_R^\dagger(y-t) \rangle &   \langle\lambda_L(x+t) \lambda_L^\dagger(y+t) \rangle
    \end{pmatrix}}  \\
    &
          \rule{0pt}{.8cm}
    = \begin{pmatrix}
        C(x-y)&                 e^{\ii\alpha_v}C(x+y)\\
        e^{-\ii\alpha_v}C(-x-y)&  C(-x+y)
    \end{pmatrix} \equiv \mathscr{C}(x,y;\alpha_v),
\end{split}
\end{equation}
in terms of \eqref{eq:DiracCorr} and of the phase $\alpha_v$ characterising the boundary condition (\ref{eq:vectorCond}).

\paragraph{Axial phase:}
In this case it is convenient to denote the massless Dirac field on the half line $x\geqslant 0$ as follows
\begin{equation}
    X(x,t) ={ \begin{pmatrix}
        \chi_R^\dagger(x-t)\\
        \chi_L(x+t)
    \end{pmatrix},}
\end{equation}
where $\chi_{R}$ and $\chi_{L}$ are the two chiral components. 
The global energy conservation condition is solved also by the following boundary condition at $x=0$
\begin{equation}
\label{eq:axialCond}
    \chi_{R}(t) = e^{-\ii \alpha_a} \chi^\dagger_{L}(t),\qquad \alpha_a \in [0, 2\pi)\,.
\end{equation}
This boundary condition preserves the axial symmetry (\ref{eq:axialSymm}) and breaks the vector one (\ref{eq:vectorSymm}).

In terms of the doublet $X(x,t)$, 
the correlation matrix turns out to be identical to \eqref{eq:vectorCorr} for the vector phase \cite{mt-21}
\begin{equation}
\label{eq:axialCorr}
    \langle X(x,t) X^\dagger(y,t) \rangle
    =
    \begin{pmatrix}
        \langle\chi_R^\dagger(x-t) \chi_R(y-t) \rangle&   \langle\chi_R^\dagger(x-t) \chi_L^\dagger(y+t) \rangle
        \\
        \rule{0pt}{.5cm}
        \langle\chi_L(x+t) \chi_R(y-t) \rangle&           \langle\chi_L(x+t) \chi_L^\dagger(y+t) \rangle
    \end{pmatrix} 
    = \mathscr{C}(x,y;\alpha_a)\,,
\end{equation}
in terms of the phase $\alpha_a$ which parameterises the family of boundary conditions (\ref{eq:axialCond}).

Since the correlation matrices \eqref{eq:vectorCorr} and \eqref{eq:axialCorr} are the same in the vector and in the axial phase, 
one can treat them in a unified way by introducing the doublet
\begin{equation}\label{eq:bothPhases}
    \Psi(x,t) = \begin{pmatrix}
        \psi_R(x-t)\\
        \psi_L(x+t)
    \end{pmatrix} = \begin{cases}
        \Lambda(x,t),   &\text{vector phase}\\
        X(x,t),      &\text{axial phase}\,,
    \end{cases} 
\end{equation}
which will be used throughout this manuscript. 

\section{Entanglement Hamiltonian for the union of disjoint intervals}\label{sec:EH}

In this section we study the entanglement Hamiltonians of a subsystem 
made by the union of an arbitrary number of intervals 
for the massless Dirac field on the half line,
extending the result obtained in \cite{mt-21} for a single interval. 
In order to carry out this analysis, we find it worth reviewing first the
analogous result for the massless Dirac field on the line,
performed in \cite{ch-09, achp-18}.

\subsection{Entanglement Hamiltonian on the line}

The Bisognano-Wichmann theorem \eqref{eq:BWth} does not hold in several cases, 
like e.g. when the subsystem $A$ is the union of disjoint components and 
the entire system is in the ground state \cite{ch-09, achp-18,hollands,EH-disjoint2017}.
However, in some free theories the Wick's theorem allows to establish a relation between the entanglement Hamiltonian $\hat{K}_A$ and the two-point correlation function restricted to the subsystem $A$.
This is the case for the massless Dirac field on the line
where we denote by $C_A(x,y) = \langle\Psi^\dagger(x) \Psi(y) \rangle\big|_{x,y \in A}$ 
the two-point correlator restricted to $A$,
being $ A = [a_1, b_1] \cup [a_2, b_2] \cup \ldots \cup [a_n, b_n]$
the union of $n$ disjoint intervals. 
For this quadratic model, the entanglement Hamiltonian reads
\begin{equation}\label{eq:KA}
    \hat{K}_A = \frac{1}{2\pi}\int_A dx\, \Psi^\dagger(x) H_A(x,y) \Psi(y),
\end{equation}
where $H_A(x,y)$ is a quadratic kernel that we investigate in the following. 
The two-point correlation function restricted to $A$ can then be written in terms of this kernel as \cite{pkl-99, Peschel2009, Chung2001, Peschel2003, Peschel2004, Peschel2012}
\begin{equation}\label{eq:peschel}
    C_A = \frac{1}{1+e^{H_A}}.
\end{equation}
This formula implies that $C_A$ and the kernel $H_A$ share the same eigenfunctions, 
and that the eigenvalues $e_s$ of the latter are related to the eigenvalues $\sigma_s$ of the former through
\begin{equation}\label{eq:lattPeschel}
    e_s \,=\, \log\! \left( \frac{1-\sigma_s}{\sigma_s}\right).
\end{equation}
Therefore, by finding the eigenfunctions and the eigenvalues of $C_A$ it is possible to apply this relation to obtain an analytical expression for the entanglement Hamiltonian.

For the sake of simplicity, we describe the procedure for the right-moving chiral fermions $\psi_R(x-t)$, but similar steps apply to the left-moving ones as well. 
The correlation function for $\psi_R(x-t)$ on the line is $C_A(x-y)$, given by \eqref{eq:DiracCorr}. 
Exploiting the results found in \cite{musk-book}, the spectral problem for the correlator of a chiral component of the Dirac fermions 
has been solved \cite{ch-09, achp-18}:
the eigenvalues take values in $[0,1]$ and they can be written as
\begin{equation} \label{eq:eigenvaluesCorr}
    \sigma_s = \frac{1}{2}\, \big[1 + \tanh\left ( \pi s \right )\big],\qquad s \in \RR.
\end{equation}
Each eigenvalue $\sigma_s$ has an $n$-fold degeneracy and the corresponding eigenfunctions take the form 
\begin{equation}\label{eq:eigenvectors}
    \phi_p^s(x) = k_p(x) e^{-\ii s w(x)}, \quad  w(x) =\log \! \left [ - \prod_{i=1}^n \frac{(x-a_i)}{(x-b_i)} \right ],
\end{equation}
where
\begin{equation}\label{eq:kappa}
	k_p(x) 
	= \frac{(-1)^{n+1}}{N_p} \frac{1}{(x-a_p)} \sqrt{- \prod_{i=1}^n\frac{(x-a_i)}{(x-b_i)}}\,, 
	\qquad
	N_p = \sqrt{2\pi} \sqrt{\frac{\prod_{i\neq p} (a_i - a_p)}{\prod_{i=1}^n (b_i - a_p)}}\, .
\end{equation}
These eigenfunctions form a complete orthonormal basis. 
From \eqref{eq:peschel} and (\ref{eq:eigenvaluesCorr}), one finds that
the eigenvalues of the entanglement Hamiltonian kernel $H_A$ (for right-moving chiral fermions) are
\begin{equation}\label{eq:entangEigenvaluesLeft}
    e_s = - 2\pi s.
\end{equation}
Then, the kernel 
can be written in spectral representation as follows \cite{ch-09, achp-18}
\begin{equation}\begin{split}
	H_A(x,y)=& \sum_{p = 1}^n \int_{-\infty}^{+\infty} ds\, \phi_p^s(x) \big(\! -\! 2\pi s \big) \phi_p^{s*}(y) = - k(x,y) \int_{-\infty}^{+\infty} ds\, s\, e^{-\ii s[w(x)-w(y)]},
\end{split}\end{equation}
where we have introduced 
\begin{equation}
k(x,y) \equiv 2 \pi \sum_{p = 1}^n k_p(x) k_p(y) .
\end{equation}

The integral over the eigenvalue $s$ is proportional to the derivative of a delta function
\begin{equation}\label{eq:HA}
    \int_{-\infty}^{+\infty} ds\, s\, e^{-\ii s[w(x)-w(y)]} = 2\pi \ii\,  \frac{1}{2} \left [ \frac{\partial_x}{w'(x)} - \frac{\partial_y}{w'(y)} \right ] \delta(w(x)-w(y)),
\end{equation}
which imposes that the kernel couples conjugate points $x$ and $y$
where the function $w$ takes the same value, i.e. satisfying
\begin{equation}\label{eq:w}
   w(y) = w(x). 
\end{equation}
This equation has the trivial solution $y=x$, which gives rise to a local term in the entanglement Hamiltonian, 
and also other $n-1$ non-trivial solutions $y = \tilde{x}_{p}$ with $1\leqslant p \leqslant n-1$,
which are responsible of the $n-1$ bi-local terms  in the entanglement Hamiltonian.
Using the properties of the delta function in \eqref{eq:HA},
the kernel is
\begin{equation}\begin{split}\label{eq:EH}
    H_A(x,y) =&
     \;- 2\pi k(x,y) \left [ \frac{\ii}{2} \frac{\left ( \partial_x - \partial_y \right ) \delta(x-y)}{w'(x) w'(y)}
      - \frac{\ii}{2} \sum_{p=1}^{n-1} \left( \frac{\partial_y \delta(y - \tilde{x}_p)}{w'(\tilde{x}_p)w'(y)} - \frac{\partial_x \delta(x - \tilde{y}_p)}{w'(x)w'(\tilde{y}_p)} \right) \right ] 
    \\
	=& 
	\;H_A^\text{loc}(x,y) + H_A^\text{bi-loc}(x,y).
\end{split}\end{equation}
In order to find the explicit expression of the entanglement Hamiltonian, it is convenient to use the properties of the $k_p(x)$ functions, which lead to \cite{achp-18, mt-21}
\begin{equation}\label{eq:kappap}
    k(x,x) = 2\pi \sum_{p=1}^n k_p(x) k_p(x) = w'(x), \quad
    k(x,\tilde{x}_p) = 0, \quad 
    \partial_y k(x,y)\big |_{y = \tilde{x}_p} = \frac{w'(x)}{x - \tilde{x}_p}.
\end{equation}
Plugging $H_A^\text{loc}(x,y)$ into \eqref{eq:KA} and integrating by parts, the local term yields \cite{ch-09, achp-18}
\begin{equation}\begin{split}\label{eq:CHloc}
	\hat{K}_A^\text{loc} =
	& - \int_A dx \int_A dy\, \frac{k(x,y)}{w'(x)w'(y)} \big[ \left ( \partial_x - \partial_y \right ) \delta(x-y) \big] \;\frac{\ii}{2} :\!\psi_{R}^\dagger(x) \psi_{R}(y)\!: 
	 \\
	=
	& \int_A dx \int_A dy\, \frac{k(x,y)}{w'(x)w'(y)} \delta(x-y) \; { \frac{\ii}{2} : \!\! \left ( \partial_x \psi_{R}^\dagger(x) \psi_{R}(y) - \psi_{R}^\dagger(x) \partial_y \psi_{R}(y)\right )\!\! : }
	\\
	=
	& \int_A dx\, \frac{T_R(x, t=0)}{w'(x)} \, \equiv \int_A dx\, \beta^\text{loc}(x)\, T_R(x, t=0),
\end{split}\end{equation}
where we used $\partial_x k(x,y) \big |_{y = x} = \partial_y k(x,y) \big |_{y = x}$ 
and $: \cdots :$ indicates that the corresponding operators are normal ordered. 
We recognise that the local term is the integral over $A$ of the chiral stress-energy tensor
\begin{equation} 
    T_{R}(x, t) = \frac{\ii}{2} : \!\! \left [ \partial_x \psi_{R}^\dagger(x-t) \psi_{R}(x-t) - \psi_{R}^\dagger(x-t) \partial_x \psi_{R}(x-t)\right ] \!\! :\,,
\end{equation}
weighted by the \emph{local effective inverse temperature}
\begin{equation}\label{eq:betalocgen}
    \beta^\text{loc}(x) \equiv \frac{1}{w'(x)}.
\end{equation}
When $A$ consists of one single interval, this local term agrees with the general CFT result \cite{ct}
\begin{equation}
    w(x) \,=\,  \log \!\left( \frac{b-x}{x-a} \right).
\end{equation}
This function is the uniformising transformation which maps the worldsheet into the annulus configuration and $\beta^\text{loc}(x) = 1/w'(x)$ is the predicted value for the local inverse temperature. 

The bi-local term  takes the form \cite{ch-09, achp-18}
\begin{equation}
\begin{split}\label{eq:CHbiloc}
    &\hat{K}_A^{\text{bi-loc}} = - \int_A dx \int_A dy \, 
    \big(\! - \!k(x,y) \big) \sum_{p=1}^{n-1} \left [ \frac{\partial_y \delta(y - \tilde{x}_p)}{w'(\tilde{x}_p)w'(y)} - \frac{\partial_x \delta(x - \tilde{y}_p)}{w'(x)w'(\tilde{y}_p)} \right ] \frac{\ii}{2} :\!  \psi_{R}^\dagger(x) \psi_{R}(y)\! : 
    \\
	=& 
	\; - \sum_{p=1}^{n-1} \int_A dx \int_A dy \frac{\partial_y k(x,y)}{w'(\tilde{x}_p) w'(y)} \, \delta(y - \tilde{x}_p) \;
	\frac{\ii}{2} :\! \!\left ( \psi_{R}^\dagger(x) \psi_{R}(y) - \psi_{R}^\dagger(y) \psi_{R}(x) \right )\!\!: 
	\\
	=& \; \sum_{p=1}^{n-1} \int_A dx \, \frac{1}{x-\tilde{x}_p} \, \frac{1}{w'(\tilde{x}_p)} \; T_{R}^\text{bi-loc}(x, \tilde{x}_p, t=0) \;
	= \,\sum_{p=1}^{n-1} \int_A dx\, \frac{\beta^\text{loc}(\tilde{x}_p)}{x-\tilde{x}_p}\; T_{R}^\text{bi-loc}(x, \tilde{x}_p, t=0),
\end{split}\end{equation}
where we have introduced the following \emph{bi-local operator}
\begin{equation}\label{eq:qlocdiagLeft}
    T_{R}^\text{bi-loc}(x, y, t) \,= \, \frac{\ii}{2} :\!\! \left [ \psi^\dagger_{R}(x-t) \psi_{R}(y-t) - \psi_{R}^\dagger(y-t) \psi_{R}(x-t) \right ] \!\!:\, ,
\end{equation}
that must be evaluated at the non-trivial solutions $y = \tilde{x}_p$, 
and the weight function in the integrand is
\begin{equation}\label{eq:bilocalweight}
    \beta^\text{bi-loc}(x) \equiv \frac{\beta^\text{loc}(\tilde{x}_p)}{x-\tilde{x}_p}.
\end{equation}

In order to extend this argument to the left-moving fermions, recall that the two-point correlation function \eqref{eq:DiracCorr} is equal to $C_A(-x+y)$. As a consequence, we can see that for left-moving fermions the eigenvectors of the two-point correlation matrix are again given by \eqref{eq:eigenvectors}, \eqref{eq:kappa}, while its eigenvalues are $\frac{1}{2}\left [ 1 - \tanh(\pi s) \right ]$. 
Using \eqref{eq:peschel},
we find that the entanglement spectrum for left-movers is $+2\pi s$, that is, the opposite of that of right-movers. Therefore, the previous calculations are analogous up to an additional minus sign. Summing up the contributions of the two chiral fermions, the entanglement Hamiltonian of a free massless Dirac fermion is \cite{ch-09, achp-18}
\begin{equation}\label{eq:EHtot}
    \hat{K}_A = \int_A dx\, \beta^\text{loc}(x)\, T_{00}(x) + \sum_{p=1}^{n-1} \int_A dx\, \frac{\beta^\text{loc}(\tilde{x}_p)}{x-\tilde{x}_p}\, T^\text{bi-loc}_\text{diag}(x, \tilde{x}_p, t=0),
\end{equation}
where $T_{00}$ is the energy density
\begin{equation}\label{eq:energydensity} 
\begin{split}
    T_{00}(x,t) = \frac{\ii}{2} & : \!\! \Big[ \left ( \partial_x \psi_R^\dagger(x-t) \psi_R(x-t) - \psi_R^\dagger(x-t) \partial_x \psi_R(x-t) \right ) \\
    &  \quad - \left ( \partial_x \psi_L^\dagger(x+t) \psi_L(x+t) - \psi_L^\dagger(x+t) \partial_x \psi_L(x+t)\right )  \Big]\!\!:\, ,
\end{split} 
\end{equation}
and the bi-local operator for both chiralities takes the form \cite{ch-09, achp-18}
\begin{equation}\label{eq:qlocdiag}
\begin{split}
    T^\text{bi-loc}_\text{diag}(x, y, t) = \frac{\ii}{2} &:\!\!\left [ \left ( \psi^\dagger_R(x-t) \psi_R(y-t) - \psi_R^\dagger(y-t) \psi_R(x-t) \right ) \right . \\
    &\left . \quad - \left ( \psi^\dagger_L(x+t) \psi_L(y+t) - \psi_L^\dagger(y+t) \psi_L(x+t) \right ) \right ]\!\!: .
\end{split}
\end{equation}

We point out an interesting feature of the entanglement Hamiltonian \eqref{eq:EHtot}: the local term can be obtained as the limit of the bi-local one when $y \to x$.
In order to see this, let us  observe that the first order term of the Taylor expansion of the bi-local operator \eqref{eq:qlocdiag} around $y = x$ is proportional to the energy density
\begin{equation}
    T^\text{bi-loc}_\text{diag}(x, y, t) \approx - (y-x)\, T_{00}(x, t).
\end{equation}
Using this expansion, we immediately find that the limit of the bi-local term in which the conjugate point approaches $x$ reduces to the local contribution as
\begin{equation}
    \lim_{y \rightarrow x} \int_A dx\, \frac{\beta^\text{loc}(y)}{x-y}\, T^\text{bi-loc}_\text{diag}(x, y, t = 0) = \int_A dx\,\beta^\text{loc}(x)\, T_{00}(x, t=0).
\end{equation}
Beside providing a check for the consistency of \eqref{eq:EHtot}, this result suggests that the local term can be interpreted as the analogous of the bi-local one relative to the trivial solution $y = x$ of \eqref{eq:w}.

\subsection{Entanglement Hamiltonians on the half-line}

In the following we study the entanglement Hamiltonian of $n$ disjoint intervals 
$A= [a_1, b_1] \cup \ldots \cup [a_n, b_n]$ on the half line $x\geqslant 0$ with $a_1 >0$ 
for the massless Dirac field in the phases discussed in section \ref{sec:reviewDirac},
which are characterised by the boundary conditions (\ref{eq:vectorCond}) and (\ref{eq:axialCond})
and whose correlation matrices $\mathscr{C}_A (x,y; \alpha)$ are \eqref{eq:vectorCorr} and \eqref{eq:axialCorr} respectively. 
This extends the recent analysis performed in \cite{mt-21} for $n=1$.

\begin{figure}[t!]
    \includegraphics[width=1.\textwidth]{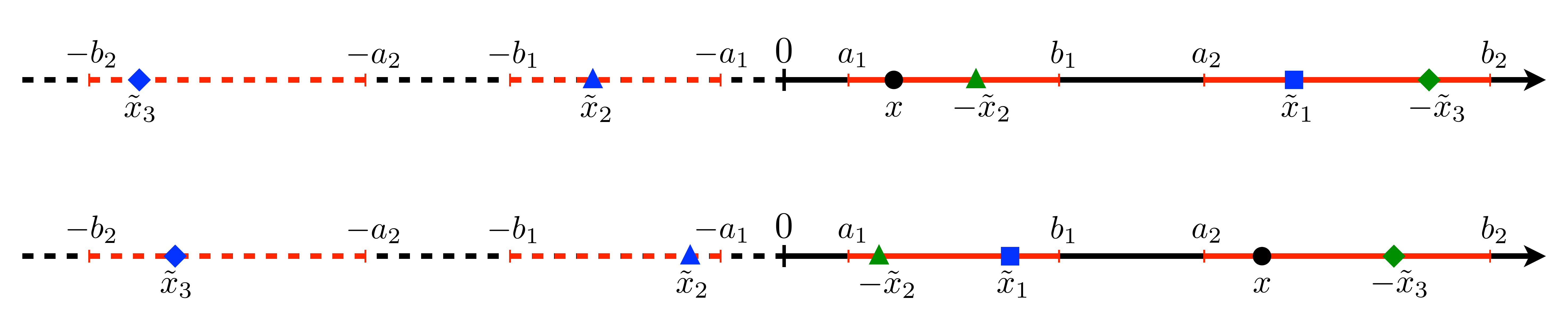}
    \vspace{-.8cm}
    \caption{Symmetric auxiliary configuration $A_\text{sym}$ on the line for the case of two intervals $A=[a_1, b_1]\cup [a_2, b_2]$ on the half line $x \geqslant 0$
    not adjacent to the boundary at $x=0$. 
    The conjugate points and their reflections are indicated for either $x\in [a_1, b_1]$ (top) or $x\in [a_2, b_2]$ (bottom). 
    The blue symbols denote the points conjugate to $x$ in $A_\text{sym}$ and the green ones the corresponding reflected points that occur in the entanglement Hamiltonian of $A$.
    }
    \label{fig:aux_geometry}
\end{figure}

\begin{figure}[t!]
\vspace{.7cm}
     \includegraphics[width=1.\textwidth]{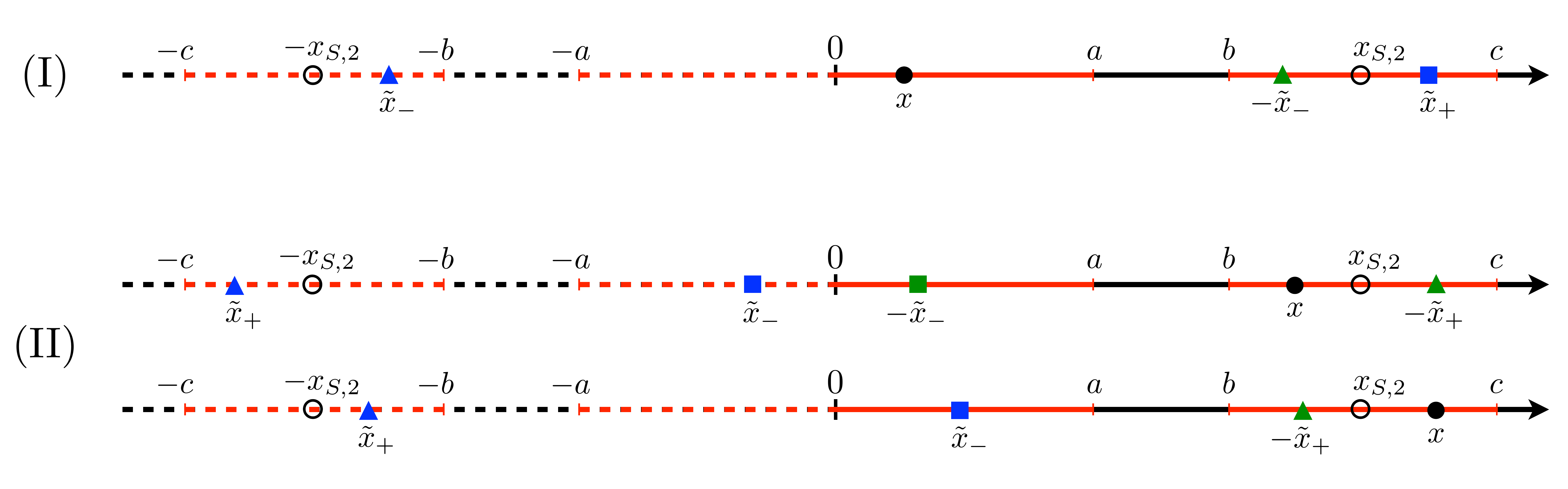}
    \vspace{-.8cm}
    \caption{Symmetric auxiliary configuration $A_\text{sym}$ on the line for the case of two intervals $A=[0, a]\cup [b,c]$ on the half line $x \geqslant 0$.
    For a given $x$, two conjugate points occur in the entanglement Hamiltonian, which are indicated following the notation established in Figure\,\ref{fig:aux_geometry}.
    The two possible cases (I) $x\in [0,a]$ and (II) $x\in [b,c]$ are represented. 
    In (II), it is distinguished whether $x\in [b, x_{S,2})$ (top) or $x\in (x_{S,2}, c]$ (bottom).
    The self-conjugate point $x_{S,2}$ is defined in \eqref{eq:selfpoint}.
    }
    \label{fig:aux_geometryAdj}
\end{figure}

In order to find the entanglement Hamiltonian we first need to solve the spectral problem 
associated to the restricted correlation matrix $\mathscr{C}_A (x,y; \alpha)$
\begin{equation}\label{eq:sp-bdy}
    \int_A dy\, \mathscr{C}_A (x,y; \alpha) \, \Phi_p^s(y) = \sigma_s\, \Phi_p^s(x).
\end{equation}
For this purpose, following \cite{mt-21},
let us consider the symmetric auxiliary configuration on the line 
$A_\text{sym} \equiv [-b_n, -a_n] \cup \ldots \cup [-b_1, -a_1] \cup [a_1, b_1] \cup \ldots \cup [a_n, b_n] \subset \mathbb{R}$,
obtained by reflecting the subsystem $A$ with respect to the position of the boundary at $x=0$ (see Fig.~\ref{fig:aux_geometry}),
and the corresponding eigenfunctions 
$\phi_p^{s\text{ (sym)}}(x) = k_p^\text{sym}(x) e^{- \ii s w_\text{sym}(x)}$  of the correlator $C_{A_\text{sym}}$ 
restricted to $A_\text{sym}$.
The number of disjoint intervals in the symmetric auxiliary geometry $A_\text{sym}$ depends on the fact if the first interval is adjacent to the boundary or not. In the former case, as depicted in Fig.~\ref{fig:aux_geometryAdj}, there is one interval which crosses the boundary and $A_\text{sym}$ is composed of $2 n - 1$ intervals. For $a_1 > 0$, shown in Fig.~\ref{fig:aux_geometry}, instead, $A_\text{sym}$ contains $2 n$ intervals since none of them crosses the boundary. For simplicity, in the following we will call $\widetilde{n} + 1$ the number of intervals contained in the symmetric auxiliary geometry, i.e., $\widetilde{n} = 2n - 1$ for $a_1 > 0$ and $\widetilde{n} = 2n - 2$ for $a_1 = 0$. 

A straightforward extension of the observation made in \cite{mt-21} leads us to 
write the eigenfunctions of the spectral problem \eqref{eq:sp-bdy} as follows
\begin{equation} \label{eq:eigenvectorBound}
    \Phi_p^s(x) = \begin{pmatrix}
        e^{\ii \alpha} \,\phi_p^{s\text{ (sym)}}(x)
        \\
        \rule{0pt}{.6cm}
        \phi_p^{s\text{ (sym)}}(-x)
    \end{pmatrix},
    \qquad
    1 \leq p \leq \widetilde{n} + 1\,,
\end{equation}
whose corresponding eigenvalues are $\sigma_s = \frac{1}{2}\left [1+\tanh(\pi s)\right ]$, with $s\in \mathbb{R}$. This solution of the spectral problem on the half line allows us to write the entanglement Hamiltonian kernel in (\ref{eq:KA})
through  its spectral representation as follows
\begin{equation}\begin{split}\label{eq:W}
	H_{A} (x,y) 
	=
	& \sum_{p=1}^{\widetilde{n} + 1} \int_{-\infty}^{+\infty} ds\, \Phi_p^s(x) \big(\! -\! 2\pi s \big) \Phi_p^{s\dagger}(y)  \\
	=
	& \sum_{p=1}^{\widetilde{n} + 1} \int_{-\infty}^{+\infty} ds\, \big(\! -\! 2\pi s \big) \, \begin{pmatrix}
		\phi_p^s(x) \phi_p^{s*}(y)&		e^{\ii \alpha}\phi_p^s(x) \phi_p^{s*}(-y)\\
		 \rule{0pt}{.5cm}
		e^{-\ii \alpha}\phi_p^s(-x) \phi_p^{s*}(y)&	\phi_p^s(-x) \phi_p^{s*}(-y)
	\end{pmatrix}  
	\\
	 \rule{0pt}{.9cm}
	=& \begin{pmatrix}
		- k^\text{sym}(x, y) W_\text{sym}(x, y)&			        - e^{\ii\alpha} k^\text{sym}(x,-y) W_\text{sym}(x, -y)\\
		 \rule{0pt}{.5cm}
		- e^{-\ii\alpha} k^\text{sym}(x, -y) W_\text{sym}(-x, y)& 	- k^\text{sym}(x, y) W_\text{sym}(-x, -y)
    \end{pmatrix},
\end{split}\end{equation}
where $x,y \in A$,
we used that $k^\text{sym}(x, -y) = k^\text{sym}(-x, y)$ 
for $A_\text{sym}$ and, for $x, y \in A$, we have introduced
\begin{equation}\begin{split}
    W_\text{sym}(x, y) \equiv \int_{-\infty}^{+\infty} ds\, s\, e^{-\ii s [w_\text{sym}(x)-w_\text{sym}(y)]} 
    = \pi \ii \left [ \frac{\partial_x}{w_\text{sym}'(x)} - \frac{\partial_y}{w_\text{sym}'(y)} \right ] \delta(w_\text{sym}(x)-w_\text{sym}(y)).
\end{split}\end{equation}

Unlike the case without boundaries, 
now the entanglement Hamiltonian is non-diagonal in the chiral fermions, as already found in \cite{mt-21} for $n=1$,
where the bi-local term was provided entirely  by the out-of-diagonal term in the kernel.
Instead, for $n > 1$ we will show that both the diagonal and out-of-diagonal elements contribute to the bi-local 
part of the entanglement Hamiltonian. 

\paragraph{Diagonal terms:}
The diagonal elements of $H_A$
are analogous to the kernel in the problem without boundary since they are localised along the solutions of the equation $w_\text{sym} (y) = w_\text{sym} (x)$.
For a subsystem $A$ with $n$ intervals, beside the trivial solution $y = x$, the equation (\ref{eq:w}) has $\widetilde{n}$ non-trivial ones $y = \tilde{x}_p$, i.e., either $2n-2$ or $2n-1$ depending if the first interval is adjacent to the boundary or not.
However, differently from the case without boundary, now the only acceptable solutions are those that belong to the original subsystem $A$; hence we only need to keep $x,y > 0$. 
While this condition is always verified for the trivial solution, for the non-trivial ones we must restrict ourselves to $\tilde{x}_p>0$ only. 
For $n=2$, in Fig.~\ref{fig:aux_geometry} and Fig.~\ref{fig:aux_geometryAdj} we show qualitatively the configurations of these points
when either $a_1 >0$ or $a_1=0$ respectively. 
In the latter case (I) and (II) correspond respectively to  $x\in A_1$ and $x\in A_2$.

Thus, the diagonal part of the kernel can be written as
\begin{equation}\begin{split}\label{eq:kernel}
   H_A^{\text{diag}}(x,y) =& -2\pi k^\text{sym}(x,y) \begin{pmatrix}
		1&	0\\
		0&	-1
	\end{pmatrix} \frac{\ii}{2} \,\bigg\{
	 \frac{\left ( \partial_x - \partial_y \right ) \delta(x-y)}{w_\text{sym}'(x) w_\text{sym}'(y)}  \\
	 \rule{0pt}{.7cm}
	&\quad
	\;\;\;\; 
	-  \sum_{p=1}^{\widetilde{n}} 
	\left [ \frac{\partial_y \delta(y - \tilde{x}_p)}{w_\text{sym}'(\tilde{x}_p)w_\text{sym}'(y)} - \frac{\partial_x \delta(x - \tilde{y}_p)}{w_\text{sym}'(x)w_\text{sym}'(\tilde{y}_p)} \right ] 
	\! \bigg\}
	\\
		 \rule{0pt}{.6cm}
	=& \; H_A^{\text{loc}}(x,y) + H_A^{\text{bi-loc, diag}}(x,y)\, ,
\end{split}\end{equation}
where we have separated the local part from the bi-local one.

The final step of the calculation is to substitute the kernel \eqref{eq:kernel} in the expression of the entanglement Hamiltonian \eqref{eq:KA} and integrate by parts, keeping in mind that now the kernel is a $2\times 2$ matrix. 
For the local component we find
\begin{equation}\begin{split}
	K_{A}^\text{loc} =&\,  \frac{1}{2\pi}\int_{A} dx \int_{A} dy\, 
	 :\!\! \Big ( \psi_R^\dagger(x), \psi_L^\dagger(x) \Big ) H_A^{\text{loc}}(x,y) \begin{pmatrix}
		\psi_R(y)\\
		\psi_L(y)
	\end{pmatrix} \!\! : 
	\\
		\rule{0pt}{.7cm}
	=& \int_{A} dx \int_{A} dy \; \frac{k(x,y)}{w_\text{sym}'(x)w_\text{sym}'(y)} \; \delta(x-y) 
	\\
	& \;\;\;\;\; \times  \frac{\ii}{2}
	:\!\! \left [\left (\partial_x \psi_R^\dagger(x) \psi_R(y) - \psi_R^\dagger(x) \partial_y \psi_R(y) \right ) 
	- \left (\partial_x \psi_L^\dagger(x) \psi_L(y) - \psi_L^\dagger(x) \partial_y \psi_L(y) \right ) \right ]\!\! :\,
	\\
	\rule{0pt}{.8cm}
	=& \int_{A} dx\, \frac{T_{00}(x,t = 0) }{w_\text{sym}'(x)}  = \int_{A} dx\, \beta^\text{loc}_\text{sym}(x)\, T_{00}(x, t = 0)\, ,
\end{split}\end{equation}
where we recognise the same form of the local term occurring in the case without boundaries in \eqref{eq:CHloc}. 
However, now the local inverse temperature corresponds to the one for the auxiliary configuration $A_\text{sym}$
\begin{equation}
    \beta^\text{loc}_\text{sym}(x) = \frac{1}{w_\text{sym}'(x)}.
\end{equation}
Similarly, by plugging \eqref{eq:kernel} in \eqref{eq:KA},
for the diagonal bi-local component we find 
\begin{equation}\label{eq:diagBiloc}\begin{split}
K_{A}^\text{bi-loc, diag} =&\, \frac{1}{2\pi} \int_{A} dx\, \int_{A} dy\,   : \!\! {\Big ( \psi_R^\dagger(x), \psi_L^\dagger(x) \Big ) H_A^{\text{bi-loc, diag}}(x,y)\begin{pmatrix}
		\psi_R(y)\\
		\psi_L(y)
	\end{pmatrix}} \!\! : 
	\\
		\rule{0pt}{.8cm}
	=& 
	\;- \sum_{p=1}^{\widetilde{n}} \int_{A} dx \int_{A} dy\;  \frac{\partial_y k^\text{sym}(x,y)}{w_\text{sym}'(\tilde{x}_p) w_\text{sym}'(y)} \; \delta(y-\tilde{x}_p) 
	\\ 
	&\times {\frac{\ii}{2} :\!\! \left [\left (\psi_R^\dagger(x)\psi_R(y) - \psi_R^\dagger(y)\psi_R(x) \right ) - \left ( \psi_L^\dagger(x)\psi_L(y) - \psi_L^\dagger(y)\psi_L(x)\right ) \right ]\!\! :}\, 
	\\
			\rule{0pt}{.8cm}
	=& \;\sum_{p=1}^{\widetilde{n}} \int_{A} dx \left( \frac{1}{x-\tilde{x}_p} \; \frac{1}{w_\text{sym}'(\tilde{x}_p)} \right)  \Theta(\tilde{x}_p)\, T^\text{bi-loc}_\text{diag}(x,\tilde{x}_p, t = 0) 
	\\
			\rule{0pt}{.8cm}
	=& \; \sum_{p=1}^{\widetilde{n}} \int_{A} dx\, \frac{\beta^\text{loc}_\text{sym}(\tilde{x}_p)}{x-\tilde{x}_p}\, \Theta(\tilde{x}_p)\,T^\text{bi-loc}_\text{diag}(x,\tilde{x}_p, t=0),
\end{split}\end{equation}
where we recognise the same bi-local operator \eqref{eq:qlocdiag}, 
which occurs in the bi-local term of the case without boundary
and does not mix different chiralities. 
Notice that the integration of the Dirac delta on the finite domain $A > 0$ gives rise to the Heaviside theta function $\Theta(\tilde{x}_p)$, which guarantees that the only solutions that contribute to the final result are $\tilde{x}_p>0$, i.e., those belonging to the physical subsystem.
This term does not depend on the boundary condition; hence it is the same in both phases.
The difference of the term \eqref{eq:diagBiloc} with respect to the case without boundary is the presence of the theta function $\Theta(\tilde{x}_p)$, which imposes
that the conjugate points belong to $A$.

\paragraph{Out-of-diagonal terms:}
The out-of-diagonal terms in (\ref{eq:W}) provide the coupling between the two different chiralities, 
as already observed for the single interval $A=[a_1,b_1]$ in \cite{mt-21}.
These components are proportional to the distribution $W_\text{sym}(x,-y)$.
Thus, we need to solve the equation $w_\text{sym}(y) = w_\text{sym}(-x)$ constrained by $x,y > 0$. 
Knowing the solutions to the problem $w_\text{sym}(y) = w_\text{sym}(x)$, we immediately find that these equation is satisfied by $y = -x$ and $y = -\tilde{x}_p$.
Because of the condition $x,y >0$, the solution $y = -x$ cannot be accepted, so we will only have contributions from the others. 
We can therefore write the off-diagonal components of the kernel as
\begin{equation}\begin{split}\label{eq:outofd}
    H_A^{\text{bi-loc, mix}}(x,y) =& -2\pi k^\text{sym}(x,-y)\begin{pmatrix}
		0&		e^{\ii\alpha}\\
		-e^{-\ii\alpha}&	0
	\end{pmatrix} 
	\\ & \times \frac{\ii}{2} \sum_{p=1}^{\widetilde{n}} \left [ \frac{\partial_y \delta(y+\tilde{x}_p)}{w_\text{sym}'(\tilde{x}_p) w_\text{sym}'(y)} 
	+ \frac{\partial_x \delta(x+\tilde{y}_p)}{w_\text{sym}'(x) w_\text{sym}'(\tilde{y}_p)}\right ],  \qquad x, y \in A\,.
\end{split}\end{equation}

Plugging \eqref{eq:outofd} into the expression for the entanglement Hamiltonian (\ref{eq:KA}), we finally get
\begin{equation}\label{eq:mixBiloc}\begin{split}
	K_{A}^\text{bi-loc, mix} \,= & \; \frac{1}{2\pi} \int_{A} dx\, \int_{A} dy\, 
	: \!\! {\Big ( \psi_R^\dagger(x), \psi_L^\dagger(x) \Big )H_A^{\text{bi-loc, mix}}(x,y)\begin{pmatrix}
		\psi_R(y)\\
		\psi_L(y)
	\end{pmatrix}} \! \!:  \\
	=& \;  \sum_{p=1}^{\widetilde{n}} \int_{A} dx \int_{A} dy\;  \frac{\partial_y k^\text{sym}(-x,y)}{w_\text{sym}'(\tilde{x}_p) w_\text{sym}'(y)} \delta(y+\tilde{x}_p) 
	\\ 
	& \times { \frac{\ii}{2} :\!\!\left [ e^{\ii\alpha} \left (\psi_R^\dagger(x)\psi_L(y) + \psi_R^\dagger(y)\psi_L(x) \right ) - e^{-\ii\alpha} \left ( \psi_L^\dagger(x)\psi_R(y) + \psi_L^\dagger(y)\psi_R(x)\right ) \right ]\!\!:\,} 
	\\
	=& \; \sum_{p=1}^{\widetilde{n}} \int_{A} dx \; \frac{1}{x-\tilde{x}_p} \; \frac{1}{w_\text{sym}'(\tilde{x}_p)} \; \Theta(-\tilde{x}_p)\, T^\text{bi-loc}_\text{mix}(x,-\tilde{x}_p,t=0;\alpha) 
	\\
	=& \; \sum_{p=1}^{\widetilde{n}} \int_{A} dx\, \frac{\beta^\text{loc}(\tilde{x}_p)}{x-\tilde{x}_p}\; \Theta(-\tilde{x}_p)\, T^\text{bi-loc}_\text{mix}(x,-\tilde{x}_p, t=0;\alpha)\, ,
\end{split}\end{equation}
where we used the property that for a symmetrical geometry $A_\text{sym}$ the function $w_\text{sym}(x)$ is odd (and thus its derivative is even $w_\text{sym}'(-x) = w_\text{sym}'(x)$) 
and now the integration over $A$ gives rise to $\Theta(-\tilde{x}_p)$, implying that $\tilde{x}_p$ belong the reflection of $A$ on the other side of the boundary.
In this term, we find the same bi-local operator that appeared in \cite{mt-21}
\begin{equation}\label{eq:mixBilocOp}
\begin{split}
	T^\text{bi-loc}_\text{mix}(x,y, t;\alpha) 
	= \frac{\ii}{2} :\!
	&\left [ e^{\ii\alpha} \left (\psi_R^\dagger(x-t) \psi_L(y+t) + \psi_R^\dagger(y-t) \psi_L(x+t) \right ) + \right . \\
	&\left .\quad - \,e^{-\ii\alpha}  \left (\psi_L^\dagger(x+t) \psi_R(y-t) + \psi_L^\dagger(y+t) \psi_R(x-t) \right )  \right]\!\!: .
\end{split}\end{equation}
This operator is non-diagonal in the chiral fermions and is dependent on the boundary condition. In particular, as showed in \cite{mt-21}, the explicit form of \eqref{eq:mixBilocOp} changes between the two phases. 
Using \eqref{eq:bothPhases}, in the vector phase the operator is equal to
\begin{equation}\label{eq:mixVecBilocOp}\begin{split}
	T^\text{bi-loc}_\text{mix, vec}(x,y, t; \alpha_v) = \frac{\ii}{2} :\!
	&\left [ e^{\ii\alpha_v}  \left (\lambda_R^\dagger(x-t) \lambda_L(y+t) + \lambda_R^\dagger(y-t) \lambda_L(x+t) \right )  \right . \\
	&\left .\quad - \, e^{-\ii\alpha_v} \left (\lambda_L^\dagger(x+t) \lambda_R(y-t) + \lambda_L^\dagger(y+t) \lambda_R(x-t) \right ) \right]\!\!:\,,
\end{split}\end{equation}
while in the axial phase
\begin{equation}\label{eq:mixAxBilocOp}\begin{split}
	T^\text{bi-loc}_\text{mix, ax}(x,y, t; \alpha_a) = \frac{\ii}{2} :& \Big[e^{\ii\alpha_a} \Big(\chi_R(x-t) \chi_L(y+t) + \chi_R(y-t) \chi_L(x+t) \Big) 
	\\
	& \quad - e^{-\ii\alpha_a} \left (\chi_L^\dagger(x+t) \chi_R^\dagger(y-t) + \chi_L^\dagger(y+t) \chi_R^\dagger(x-t) \right )  \Big] \!\!:.
\end{split}\end{equation}
Notice that in the axial phase the non-diagonal bi-local operator violates the conservation of electric charge. 
This is a consequence of the explicit breaking of the vector symmetry. 
Another important difference with respect to the diagonal bi-local term \eqref{eq:diagBiloc} is that $\tilde{x}_p < 0$,
i.e. it must belong to the reflection of the physical subsystem with respect to the boundary.

Putting together the terms discussed above, 
as  first major result of this manuscript,
we find the entanglement Hamiltonian of a multi-interval geometry on the half line
\begin{equation}\begin{split}\label{eq:mainb}
    \hat{K}_A =&\, K_{A}^\text{loc} + K_{A}^\text{bi-loc, diag} + K_{A}^\text{bi-loc, mix} \, = \int_A dx\, \beta^\text{loc}_\text{sym}(x) \,T_{00}(x, t = 0)\,
    \\
    & + \sum_{p=1}^{\widetilde{n}} \int_A dx\, \frac{\beta^\text{loc}_\text{sym}(\tilde{x}_p)}{x-\tilde{x}_p} \,
    \Big[ \Theta(\tilde{x}_p)\, T^\text{bi-loc}_\text{diag}(x,\tilde{x}_p, t=0) %
    + \Theta(-\tilde{x}_p)\, T^\text{bi-loc}_\text{mix}(x,-\tilde{x}_p, t=0;\alpha) \Big],
\end{split}
\end{equation}
written in terms of the operators  (\ref{eq:energydensity}) and (\ref{eq:qlocdiag}), 
which do not mix the fields with different chiralities, 
and either (\ref{eq:mixVecBilocOp}) or (\ref{eq:mixAxBilocOp})
for the vector and axial phase respectively. We remind the reader that $\widetilde{n}$ is equal to $2 n - 1$ when $a_1 > 0$ and to $2 n - 2$ when $a_1 = 0$.

Let us compare (\ref{eq:mainb}) with the entanglement Hamiltonian without boundary in \eqref{eq:EHtot} in the special case of $A=A_\text{sym}$. 
Beside the obvious difference between the integration domains, 
the weight functions and the entire local terms are the same.
The main difference is due to the bi-local operator;
indeed, in (\ref{eq:mainb}) different bi-local operators occur in the terms corresponding to different $p$'s,
depending on the sign of $\tilde{x}_p$.
In particular, when $\tilde{x}_p > 0$ the corresponding bi-local operator is (\ref{eq:qlocdiag}), 
which does not mix different chiralities, 
while the bi-local operator associated to $\tilde{x}_p < 0$ is (\ref{eq:mixBilocOp})
(that becomes either (\ref{eq:mixVecBilocOp}) or (\ref{eq:mixAxBilocOp}), depending on the phase), 
which couples fields with different chirality and explicitly depends on the boundary condition parameter.

The bi-local operator $T^\text{bi-loc}_\text{mix}$ is evaluated in $-\tilde{x}$; hence $x\in [a_j,b_j]$ for some $j$, 
we have that $\tilde{x}\in [-b_j,-a_j]$ and therefore $-\tilde{x}$ belongs to the physical subsystem $A$. 
In Fig.~\ref{fig:aux_geometry} and Fig.~\ref{fig:aux_geometryAdj} the two intervals case is considered 
(with the first interval either separated or adjacent to the boundary respectively):
the same type of marker denotes both the conjugate points belonging to the symmetric auxiliary geometry (blue symbols) 
and their reflections with respect to the boundary that belong to $A$ (green symbols).
 As discussed in \cite{mt-21}, the consequence is that in the $j$-th interval  there is one point $x_{S,j}$ that is conjugated to its own reflection, i.e. $\tilde{x}_{S,j} = -x_{S,j}$;
 hence we refer to them as \emph{self-conjugate points}. 
If $a_1>0$, in Fig.~\ref{fig:aux_geometry} these points correspond to $x$ (the black dot) that coincides with the conjugate point in the same interval. 
Since $w_\text{sym}(x)=w_\text{sym}(y)$ for conjugate points, the self-conjugate points satisfy $w_\text{sym}(x_{S,j}) = w_\text{sym}(-x_{S,j})$. 
Being $w_{\mathrm{sym}}(x)$ an odd function of $x \in A_{\mathrm{sym}}$, 
we conclude that $w_{\mathrm{sym}}(x_{S,j}) = - w_{\mathrm{sym}}(x_{S,j})$, 
which implies that the self-conjugate points are \emph{all and only} the zeroes of $w_{\mathrm{sym}}(x)$ on the positive real semi-axis.
Moreover, for a generic point $y$ conjugate to $x_{S,j}$ we have that $w_{\mathrm{sym}}(y)=w_{\mathrm{sym}}(x_{S,j})=0$, i.e. $y$ is also a self-conjugate point. 
In Fig.~\ref{fig:aux_geometry}, when the black dot coincides with a green marker in one interval, also the two remaining conjugate points coincide in the other interval.
Hence, all the self-conjugate points are conjugate among themselves. 
If $A$ is made by $n$ intervals, $w_{\mathrm{sym}}(x)=0$ admits $n$ solutions with $x\geq0$, corresponding to the self-conjugate points. 
Since $w_{\mathrm{sym}}(x)$ is a continuous monotonic function in each interval $[a_j, b_j]$, $j=1, \dots,n,$ 
which tends to $-\infty$ as $x \to a_j$ and to $+\infty$ as $x \to b_j$, 
each interval contains one and only one self-conjugate point.

Another interesting feature of our main result \eqref{eq:mainb} occurs 
when the first interval is adjacent to the boundary, i.e. for $a_1=0$, 
as depicted in Fig.~\ref{fig:aux_geometryAdj}. 
In this case, the boundary at $x=0$ trivially satisfies $w_{\mathrm{sym}}(x=0)=0$, 
meaning that it is also a self-conjugate point, albeit a degenerate one. 
Since all self-conjugate points are conjugate among themselves, 
the other $n-1$ self-conjugate points are conjugated to the boundary and $x_{S,j} \in [a_j, b_j]$, with $j=2, \dots, n$.
A cross-over in the bi-local operator occurs at these points. 
Indeed, when $x \in [a_j, b_j]$, for $x \in [a_j, x_{S,j}]$ we have $\tilde{x}_p < 0$ and therefore the bi-local operator is non-diagonal;
while for $x \in [x_{S,j}, b_j]$ we have $\tilde{x}_p > 0$ and the corresponding operator is diagonal (see also Fig.~\ref{fig:aux_geometryAdj}).
Despite this change of operator, the entanglement Hamiltonian \eqref{eq:mainb} is continuous at $x_{S,j}$;
indeed, from the boundary conditions \eqref{eq:vectorCond} and \eqref{eq:axialCond}, 
both in the vector and in the axial phase, at $x_{S,j}$ one can show that
\begin{equation}\label{eq:discontinuous}
    T^\text{bi-loc}_\text{mix}(x_{S,j}, 0, t; \alpha) = T^\text{bi-loc}_\text{diag}(x_{S,j}, 0, t),
\end{equation}
for the non-diagonal operator (\ref{eq:mixBilocOp}) and the diagonal operator (\ref{eq:qlocdiag}).
We stress that this cross-over occurs only when the first interval is adjacent to the boundary
because the origin does not belong to $A_\text{sym}$ when $a_1 > 0$.
Thus, for $n \geqslant 2$ an important qualitative difference is observed between the cases
where $a_1 > 0$ and the ones where $a_1 =0$.

In the next section we discuss the entanglement Hamiltonian of two disjoint intervals on the half line when $a_1=0$.
This is the simplest example where the cross-over in the bi-local term described above is explicitly realised.

\subsection{Entanglement Hamiltonian for two intervals with one adjacent to the boundary}\label{sec:example}

Consider the subsystem $A = [0, a] \cup [b,c]$ on the half line, made by two disjoint intervals where the first one is adjacent to the boundary.
The corresponding auxiliary symmetric configuration is $A_\text{sym} = [-c,-b] \cup [-a,a] \cup [b, c]$, 
which is made by three disjoint intervals on the line and includes the origin. 
The function $w_\text{sym}(x)$ associated to $A_\text{sym}$ reads
\begin{equation}\label{eq:wsym}
    w_\text{sym}(x) \,=\, \log \! \left[ \frac{(x+c)(x+a)(x-b)}{(x+b)(x-a)(c-x)} \right],
\end{equation}
and it provides the following local effective inverse temperature
\begin{equation}\begin{split}\label{eq:betaLocTwoBound}
	\beta^\text{loc}_\text{sym}(x)& = \frac{1}{w_\text{sym}'(x)} = \left( \frac{2a}{x^2-a^2} - \frac{2b}{x^2-b^2} - \frac{2c}{c^2- x^2} \right)^{-1}. 
\end{split}\end{equation}
The conjugate points are the solutions of $w_\text{sym}(y) = w_\text{sym}(x)$, which is a third order algebraic equation in this case. 
One solution is the trivial $y=x$, while the other two read (see Fig.~\ref{fig:aux_geometryAdj})
\begin{equation}\begin{split}\label{eq:xtildepm}
	y = 
	\tilde{x}_\pm(x) \equiv \frac{\left (x_{S,2}^2 - x_\infty^2\right ) x \pm \sqrt{\left(x_{S,2}^2 - x_\infty^2\right )^2 x^2 + 4 x_\infty^2 (x_\infty^2-x^2)(x_{S,2}^2-x^2)} }{2(x_\infty^2-x^2)}, 
\end{split}\end{equation}
where we have introduced the two points
\begin{equation}\label{eq:selfpoint}
    x_{S,2} = \sqrt{ab+bc-ca},\;\;\qquad\;\; x_\infty = \sqrt{\frac{abc}{a-b+c}}, \;\;\qquad\;\; a < x_\infty < b < x_{S,2} < c .
\end{equation}

The point $x_{S,2}$ corresponds to the self-conjugate point described in the previous subsection. 
Indeed, one can show that $\tilde{x}_+(x_{S,2}) = -x_{S,2}$.
Moreover, we have that $\tilde{x}_-(x_{S,2}) = 0$.
This tells us that at $x_{S,2}$ the bi-local operator calculated in $\tilde{x}_-$ changes its nature from mixed to diagonal, 
in agreement with the general discussion made in the previous subsection. 
The points $x_\infty$ and $-x_\infty$ are instead the only poles of respectively $\tilde{x}_+$ and $\tilde{x}_-$; because neither belong to the subsystem $A$, we conclude that as expected \eqref{eq:xtildepm} is analytical in $A$.

\begin{figure}
    \centering
    \includegraphics[width=\textwidth]{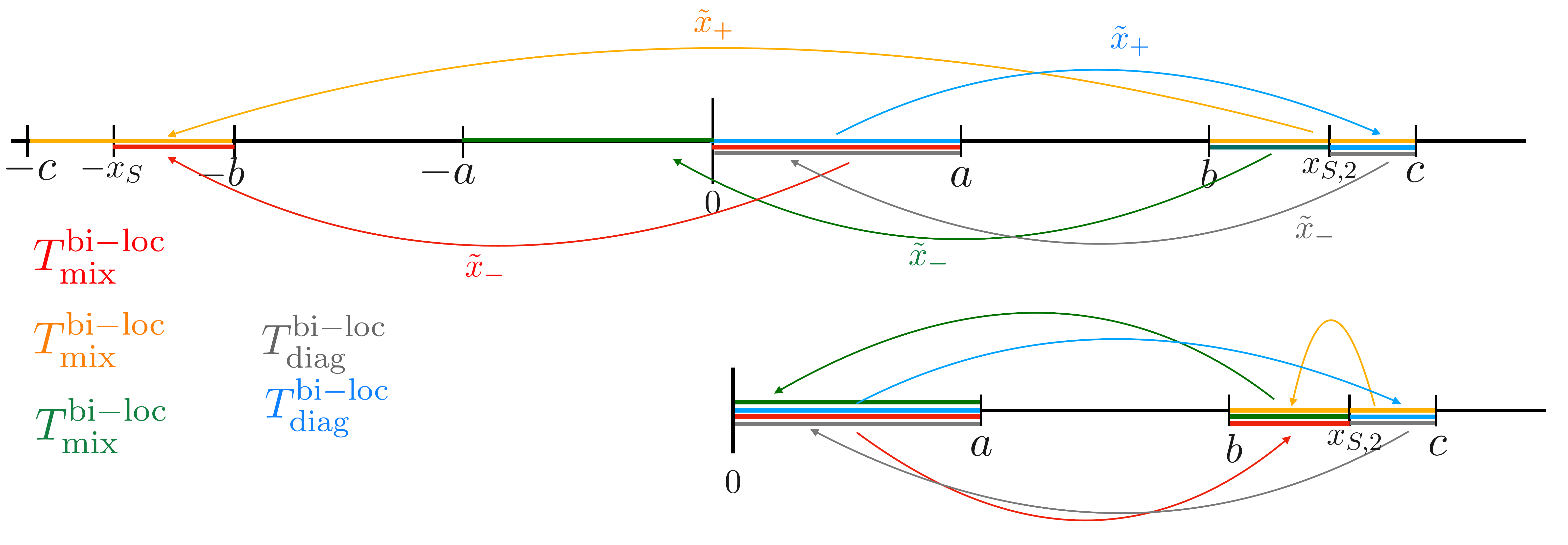}
    \caption{%
    Pictorial representation of the mappings $\tilde{x}_+$ and $\tilde{x}_-$ of two intervals with the first adjacent to the boundary, according to \eqref{eq:entTwoAttached}. 
    In the top panel, we report how the intervals $[0, a]$, $[b, x_{S,2}]$ and $[x_{S,2}, c]$ are mapped under conjugation in the auxiliary symmetric geometry $A_\text{sym}$ (see \eqref{eq:selfpoint} for the self-conjugate point $x_{S,2}$); 
    in the bottom panel, the $x<0$ points are reflected with respect to the boundary. In $A_\text{sym}$, the operators coupling points on the same side of the boundary are diagonal in the chiral fermions, 
    while the others are mixed.}
    \label{fig:entTwoAttached}
\end{figure}

Under (\ref{eq:xtildepm}), the other points of the subsystem $A$ are mapped as in Fig.~\ref{fig:entTwoAttached}
\begin{equation}\label{eq:entTwoAttached}
    \begin{cases}
        \tilde{x}_+\!\left([0,a]\right) = [x_{S,2}, c]\\
        \tilde{x}_+\!\left([b,c]\right) = [-c, -b]\,,
    \end{cases}\qquad \begin{cases}
        \tilde{x}_-\!\left([0,a]\right) = [-x_{S,2}, -b]\\
        \tilde{x}_-\!\left([b,x_{S,2}]\right) = [-a, 0]\\
        \tilde{x}_-\!\left([x_{S,2},c]\right) = [0, a]\,.
    \end{cases}
\end{equation}

Taking into account the constraints due to the theta functions,
the expression \eqref{eq:mainb} specialised to this bipartition of the half line becomes
\begin{equation}\begin{split}\label{eq:ham_two_attach}
	\hat{K}_{A} =& \int_{A} dx\, \beta^\text{loc}_\text{sym}(x)\, T_{00}(x) 
	 + \int_0^a dx\, \frac{\beta^\text{loc}_\text{sym}(\tilde{x}_+)}{x-\tilde{x}_+}\, T^\text{bi-loc}_\text{diag}(x,\tilde{x}_+, 0) 
	 \\
	&+ \int_b^c dx\, \frac{\beta^\text{loc}_\text{sym}(\tilde{x}_+)}{x-\tilde{x}_+}\, T^\text{bi-loc}_\text{mix}(x,-\tilde{x}_+, 0; \alpha) 
	+ \int_{x_{S,2}}^{c} dx\, \frac{\beta^\text{loc}_\text{sym}(\tilde{x}_-)}{x-\tilde{x}_-}\, T^\text{bi-loc}_\text{diag}(x,\tilde{x}_-, 0)
	\\ 
	&+ \left ( \int_0^a + \int_b^{x_{S,2}} \right ) dx\, \frac{\beta^\text{loc}_\text{sym}(\tilde{x}_-)}{x-\tilde{x}_-}\, T^\text{bi-loc}_\text{mix}(x,-\tilde{x}_-, 0; \alpha).
\end{split}\end{equation}
In the remaining part of this section, we study some relevant limits of this expression that provide the known results in \cite{mt-21,ch-09,achp-18}.

\paragraph{Limit $a \to 0$:}
When the length of the first interval vanishes, we expect to recover the result of \cite{mt-21} for the half line bipartite by an interval. 
For $x\in[b,c]$, the conjugate point $\tilde{x}_+$ has a finite limit and it reduces to the expression one would obtain for two symmetric intervals $[-c,-b] \cup [b,c]$
\begin{equation}
    \tilde{x}_+ \longrightarrow \tilde{x} \equiv - \frac{b c}{x} \, ,
\end{equation}
while the other conjugate point tends to zero with a linear correction in $a$, $\tilde{x}_- \sim \mathcal{O}(a)$. The self-conjugate point $x_{S,2}$ assumes the value of the one-interval case computed in \cite{mt-21}
\begin{equation}
    x_{S,2} \longrightarrow \sqrt{b c}\,.
\end{equation}
Notice that this point remains self-conjugate in the limit, i.e. $\tilde{x}(\sqrt{bc}) = -\sqrt{bc}$.

As for the local effective inverse temperature, in this limit the expression found in \cite{mt-21} is obtained
\begin{equation}\label{eq:tonnimintchev}
    \beta^\text{loc }_\text{sym}(x) \longrightarrow \frac{(x^2-b^2)(c^2-x^2)}{2(c-b)(bc+x^2)}\,.
\end{equation}
By employing also the limits of the conjugate points, we find that the weight function of the bi-local term calculated in $\tilde{x}_+$ reproduces the analogous weight in the result of \cite{mt-21}
\begin{equation}\label{eq:tonnimintchevbilocal}
    \frac{\beta^\text{loc }_\text{sym}(\tilde{x}_+) }{x-\tilde{x}_+} \longrightarrow \frac{b c (x^2 - b^2)(c^2 - x^2)}{2 (c-b) x (bc + x^2)^2}.
\end{equation}
Instead, by employing the explicit expression of $\tilde{x}_-$ in terms of the entangling points, one finds that the corresponding weight function vanishes: $\beta^\text{loc }_\text{sym}(\tilde{x}_-) \sim \mathcal{O}(a)$.

Thus, after the limits only the bi-local term mixing the chiralities (evaluated in $-\tilde{x}_+$) remains.
This finally reproduces the result obtained in \cite{mt-21}
\begin{equation}\label{eq:limit1}
	K_{A} \longrightarrow \int_b^c dx\, \beta^\text{loc }_\text{sym}(x)\, T_{00}(x) + \int_b^c dx\, \frac{\beta^\text{loc }_\text{sym}(\tilde{x})}{x-\tilde{x}}\, T^\text{bi-loc}_\text{mix}(x,-\tilde{x}, 0; \alpha).
\end{equation}

\paragraph{Limit $c \to b$:}
Let $\ell_2 \equiv c-b$ be the length of the second interval. In the limit $\ell_2 \to 0$ in which this interval vanishes, we expect to recover the result for a single interval $[0, a]$ adjacent to the
 boundary; in this case the entanglement Hamiltonian is purely local and the local effective inverse temperature is the parabolic law \eqref{eq:oneInterval} for the entanglement Hamiltonian of a single interval in a CFT. 
 For $x\in [0,a]$, in this limit both conjugate points tend to a constant $\tilde{x}_{\pm} \sim {\pm}b + \mathcal{O}(\ell_2)$ and the subleading correction is linear. 
The self-conjugate point tends to $b$ as well, as expected from the fact that the interval it belongs to vanishes.

The limit of the local effective inverse temperature reproduces the parabolic law \eqref{eq:oneInterval}
\begin{equation}\label{beta-biloc-limit-1}
    \beta^\text{loc}_\text{sym} \longrightarrow \frac{a^2-x^2}{2a},
\end{equation}
while both weight functions occurring in the bi-local terms vanish as $\beta^\text{loc}_\text{sym}(\tilde{x}_\pm) \sim \mathcal{O}(\ell_2)$.
Thus, the only term remaining after the limit is the local one and the entanglement Hamiltonian reduces to the known parabolic result \eqref{eq:oneInterval} for one interval adjacent to the boundary.

\paragraph{Limit $b \to a$:} 
Let $d \equiv b - a$ be the distance between the two intervals. In the limit $d\to 0$, they become adjacent to each other and 
the parabolic law \eqref{eq:oneInterval} for a single interval $[0,c]$ adjacent to the boundary is expected again. 
However, while in the limit $c\to b$ only the local term of the first interval contributed to the limit, now the parabolic law will be reproduced by gluing together the local inverse temperature of the two intervals $[0,a]\cup [a+d,c]$ when $d \to 0$. Indeed, for $x \in [0,c]$ the local effective inverse temperature will reduce to
\begin{equation}\label{eq:half}
    \beta^\text{loc}_\text{sym} (x) \sim \frac{c^2 - x^2}{2c} + \mathcal{O}(d).
\end{equation}
We stress that for $d$ small but finite, $\beta^\text{loc}_\text{sym}$ vanishes at the extrema $a$ and $a+d$. Only when $d = 0$ exactly these zeros disappear, giving the expected limit.

Both conjugate points tend to a constant up to a linear correction $\tilde{x}_{\pm} \sim {\pm}a + \mathcal{O}(d)$ and the weight functions of the non-local terms instead vanish linearly in $d$: $\beta^\text{loc}_\text{sym}(\tilde{x}_\pm) \sim \mathcal{O}(d)$.
Thus, for this geometry we only find a local term, whose weight function is the half-parabola in \eqref{eq:oneInterval}, \eqref{eq:half}, as one could expect for one interval at the beginning of a semi-infinite line.

\paragraph{Limit $b \to \infty$, with finite $\ell_2$:}
It is worth considering the limit of large separation between the two intervals, while their lengths are kept fixed. 
The local effective inverse temperature reduces to the parabolic law \eqref{eq:oneInterval} in the corresponding interval
\begin{equation}
    \beta^\text{loc}_\text{sym} \longrightarrow \begin{cases}
\displaystyle        \frac{a^2-x^2}{2a},        &  \hspace{.8cm} x\in [0,a]
\\
\rule{0pt}{.9cm}
\displaystyle        \frac{(x-b)(c-x)}{(c-b)},   &\hspace{.8cm}  x\in [b,c]\,.
    \end{cases}
\end{equation}
As for the weight function in the bi-local terms in (\ref{eq:ham_two_attach}), 
while both the numerators $\beta^\text{loc}_\text{sym}(\tilde{x}_+)$ and $\beta^\text{loc}_\text{sym}(\tilde{x}_-)$ remain finite, 
the denominators diverge as $x-\tilde{x}_\pm \sim \mathcal{O}(b)$. 
Thus, all the bi-local terms vanish in this limit and we obtain the sum of two local entanglement Hamiltonians, one for the single interval $[0,a]$ adjacent to the boundary and one for the single isolated interval $[b, c]$ on the full line, i.e. 
\begin{equation}
    \hat{K}_A \longrightarrow 
    \int_0^a dx\, \beta^\text{loc}_\text{sym}(x)\, T_{00}(x) + \int_b^c dx\, \beta^\text{loc}_\text{sym}(x)\,  T_{00}(x)\,.
\end{equation}
This result agrees with the intuition that the second interval is not affected by the presence of the boundary in this limit when it is very far from it.

\paragraph{Limit $\ell_2 \to \infty$, with finite $b$:} 
We find it interesting to consider the limiting regime where $\ell_2 \to \infty$ while $b$ is kept fixed. 
This limit has been explored also in  \cite{mt-21} in the special case of $a=0$.
The local effective inverse temperature reduces to 
\begin{equation}\label{eq:limBetaLoc_cToInf}
    \beta^\text{loc }_\text{sym}(x) \longrightarrow  \frac{(x^2-a^2)(x^2-b^2)}{2(b-a)(ab+x^2)} \geqslant 0,
    \qquad\;\;
    x\in A,
\end{equation}
where now the right hand side is the opposite of the local inverse temperature \eqref{eq:tonnimintchev} in the entanglement Hamiltonian 
of a single interval $[a,b]$ on the half line \cite{mt-21}.

As for the weight functions of the bi-local terms in (\ref{eq:ham_two_attach}),
for the one corresponding to $\tilde{x}_+$ we find 
\begin{equation}
    \frac{\beta^\text{loc }_\text{sym}(\tilde{x}_+) }{x-\tilde{x}_+} \longrightarrow \frac{(b^2-x^2)(x^2-a^2)}{2(b-a)x(ab-x^2)}, %
\end{equation}
which is not vanishing. 
However, since $\tilde{x}_+ \sim \mathcal{O}(\ell_2)$, i.e. it is divergent, the bi-local operators calculated in this point do not contribute because the fermionic fields $\psi_i(\tilde{x}_+)$ with $i=L,R$ vanish as $\tilde{x}_+\to \infty$, as already observed in \cite{mt-21} for a similar case. 
On the other hand, for the mapping $\tilde{x}_-$ we find
\begin{equation}
\tilde{x}_- \longrightarrow \tilde{x} = - \frac{a b}{x},
\end{equation}
and the corresponding weight function reproduces the opposite of the bi-local weight \eqref{eq:tonnimintchevbilocal} obtained for one interval $[a,b]$ on the half line\cite{mt-21}
\begin{equation}\label{eq:bilocalLimL2Inf}
    \frac{\beta^\text{loc }_\text{sym}(\tilde{x}_-) }{x-\tilde{x}_-} \longrightarrow \frac{ab (x^2 - a^2)(x^2 - b^2)}{2 (b-a) x (ab + x^2)^2}.
\end{equation}
Thus, the entanglement Hamiltonian (\ref{eq:ham_two_attach}) in this limit simplifies to 
\begin{equation}
\label{KA-limit-l2-inf}
\begin{split}
	\hat{K}_{A} \longrightarrow  \int_A dx\, \beta^\text{loc }_\text{sym}(x)\, T_{00}(x)  
	+ \int_A dx\, \frac{\beta^\text{loc}_\text{sym}(\tilde{x})}{x-\tilde{x}}\; T^\text{bi-loc}_\text{mix}(x,-\tilde{x}, 0; \alpha),
\end{split}\end{equation}
with local inverse temperature and bi-local weight function given by \eqref{eq:limBetaLoc_cToInf} and \eqref{eq:bilocalLimL2Inf}, respectively.

We remark that the entanglement Hamiltonian \eqref{KA-limit-l2-inf} is similar to the one of the single interval $[a,b]$ on the half line \cite{mt-21} (see also \eqref{eq:limit1}),
with the crucial difference that now the integration domain is the complement $A = [0,a] \cup [b, \infty )$ on the half line. 
Combining (\ref{KA-limit-l2-inf}) with the entanglement Hamiltonian found in \cite{mt-21}, 
one obtains the full modular Hamiltonian $\hat{K}_A \otimes \Id_B - \Id_A \otimes \hat{K}_B$ 
(here $\Id_X$ denotes the identity on the subsystem $X$)
for the bipartition of the half line $x\geqslant 0$ given by $[a,b]$ with $a>0$.

\subsection{Entanglement entropy}

Let us consider a bipartition of a real line (i.e. without boundaries) where the subsystem is $A_{\mathrm{sym}}=[-b_n,-a_n], \dots, [-b_1,a_1] \cup [a_1,b_1],\dots [a_n,b_n]$, with $a_1 > 0$,
which is composed of $2n$ intervals placed in a symmetric position with respect to the origin $0$, as shown in Fig.~\ref{fig:aux_geometry}. 
For the massless Dirac fermion, the R\'enyi entropies are \cite{CFH,cg-08, fps-09, cct-09, cct-11, ctt-14}
\begin{equation} \label{eq:Diracentropy}
\begin{split}
   S_{A_{\mathrm{sym}}}^{(m)} =\;\, 
   &
   \frac{m+1}{6m} \, \bigg(2\sum_{i , j}^n  \log |a_i-b_j| - 2\sum_{i<j}^n \Big[\, \log |a_i-a_j| + \log |b_i-b_j| \, \Big] -2n \log \epsilon 
   \\  
   & \hspace{.9cm}
   +\sum_{i , j}^n  \Big[ \log |b_i+b_j|+ \log |a_i+a_j| \,\Big]
   -\sum_{i\leqslant j}^n \Big[ \log |a_i+b_j| + \log |b_i+a_j| \,\Big]\bigg)  
   \\
   =\;\,& \frac{m+1}{6m} \sum_{j= 1}^{n} \left [ w_\text{sym}(b_j-\epsilon) - w_\text{sym}(a_j+\epsilon) \right ],
\end{split}\end{equation}
where $\epsilon>0$ is an ultraviolet infinitesimal cut-off. The fact that the sum in the last line only runs from $1$ to $n$, despite the presence of $2n$ intervals, is a consequence of the symmetric geometry we are considering. 
This result can be obtained by writing the moments of the reduced density matrix as the correlation functions of the branch-point twist fields \cite{cc-04} or from the knowledge of the eigenvalues and eigenvectors of the entanglement Hamiltonian \cite{ch-09}. Following the latter approach, let us compute the R\'enyi entropies when the subsystem $A$ is made by $n$ disjoint intervals on the half-line.
When $n=1$, the solution of the spectral problem for the entanglement Hamiltonian has been already used in \cite{mt-21} to write the R\'enyi entropies,
finding that they are half of the value of \eqref{eq:Diracentropy} for the auxiliary symmetric geometry $A_\text{sym} = [-b,-a] \cup [a, b]$ on the real line.
In this section, we extend this analysis to $A= \cup_{i=1}^n [a_i, b_i]$ on the half line
for $n \geqslant 1$.

The R\'enyi entanglement entropies of order $m\geqslant 2$ in terms of the correlation matrix restricted to $A$ are given by 
\cite{Peschel2009, Chung2001, Peschel2003, Peschel2004}
\begin{equation}
    S_A^{(m)} = \frac{1}{1-m} \Tr\left [ g_m\! \left ( C_A \right ) \right ], \quad \text{ with} \quad g_m\!\left (z\right ) = \log [z^m + \left ( 1 - z \right )^m].
\end{equation}
By employing the expression \eqref{eq:eigenvectorBound} of the eigenvectors $\Phi_p^s(x)$ of the correlation matrix,
of the eigenvalues \eqref{eq:eigenvaluesCorr}
and introducing  $A_\epsilon \equiv \cup_{i=1}^n [a_i + \epsilon , b_i -\epsilon] \subset A$,
one finds
\begin{equation}\label{eq:sn}
    S_A^{(m)} = \frac{1}{1-m} \sum_{p = 1}^{2n} \int_{-\infty}^{+\infty} ds \int_{A_\epsilon} dx\, g_m\!\left ( \sigma_s \right ) \Tr \left [ \Phi_p^s(x) \Phi_p^{s*}(x) \right ].
\end{equation}
From the properties of the eigenfunctions reported in \eqref{eq:kappap}, we obtain 
\begin{equation}\label{eq:trace}
   \sum_{p = 1}^{2n} \Tr \left [ \Phi_p^s(x) \Phi_p^{s*}(x) \right ]=2 \sum_{p = 1}^{2n}k_p(x)k_p(x)=\frac{w'(x)}{\pi}\,,
\end{equation}
which  is independent of the parameters $\alpha$ and $s$. 
This leads to the factorisation of the two-fold integral in \eqref{eq:sn}. 
Plugging \eqref{eq:trace} into the integral \eqref{eq:sn} and exploiting $\int_{-\infty}^{+\infty} ds\,g_m\!\left ( \sigma_s \right ) = \frac{\pi \left ( 1-m^2 \right )}{12 m}$, we finally find
\begin{equation}
    S_A^{(m)} = \frac{m+1}{12m} \sum_{j= 1}^n \big[ w_\text{sym}(b_j-\epsilon) - w_\text{sym}(a_j+\epsilon) \big],
\end{equation}
which is half of the value for the duplicated geometry without a boundary in \eqref{eq:Diracentropy}.  
This is consistent with what one would expect from the calculation of the correlation of twist fields in the presence of a boundary. 
We conclude this section by remarking that in \eqref{eq:trace} the dependence on the boundary scattering phase $\alpha$ cancels out. 
Thus, as already observed in \cite{mt-21}, while the entanglement Hamiltonian depends explicitly on the boundary condition parameter $\alpha$
through the non-diagonal operator $T^\text{bi-loc}_\text{mix}$, 
which takes the different forms \eqref{eq:mixVecBilocOp} and \eqref{eq:mixAxBilocOp} in the vector and axial phases respectively, 
the entanglement entropy is independent of the boundary condition. 
This is a typical scenario in which the entanglement Hamiltonian does contain more information with respect to the R\'enyi entropies, 
as mentioned in the introduction.

\section{Negativity Hamiltonian}\label{sec:NH}

A largely used entanglement witness for mixed states is the negativity \eqref{eq:neg_defT2}. 
This quantity is defined in terms of the partial transpose $\hat{\rho}_A^{T_2}$ of the reduced density matrix $\hat{\rho}_A$, 
where $A=A_1 \cup A_2$ and $T_2$ denotes the transposition with respect to the second interval.
Here we focus on a definition of negativity that is more suitable for free fermionic systems \cite{ssr-17,shapourian-19,ssr1-17,sr-19,ryu}, 
which has been then employed in several contexts (see e.g. \cite{ge-20,mbc-21,pbc-22,fg-22,cmc-22}).
The main reason is that, for such systems, the partial transpose $\hat{\rho}_A^{T_2}$, unlike $\hat{\rho}_A$, is not a Gaussian operator, but a sum of two non-commuting Gaussian operators \cite{ez-15,Eisert2018}. 
To avoid having to deal with non-Gaussian matrix (that is rather unpractical \cite{ctc-15,ctc-15b, ctc-15c}), a definition for the fermionic partial transpose has been introduced in \cite{ssr-17} 
which is equivalent to a partial time-reversal transformation in the fermionic coherent states. 
Let us denote the result of such partial time-reversal on $\hat{\rho}_A$ as $\hat{\rho}_A^{R_2}$, 
in order to distinguish it from the standard partial transposition $\hat{\rho}_A^{T_2}$.

To define $\hat{\rho}_A^{R_2}$, we start from 
\begin{equation}
\hat{\rho}_A=\frac{1}{Z}\int \mathrm{d}[\xi ]\mathrm{d}[\bar{\xi} ]\ee^{-\sum_{j}\bar{\xi}_j\xi_j} \bra{\{\xi_j\}}\hat{\rho}_A\ket{\{\bar{\xi}_j\}} \ket{\{\xi_j\}}\bra{\{\bar{\xi}_j\}},
\label{eq:gra}
\end{equation} 
where $\xi,\bar{\xi}$ are Grassman variables and $\ket{\xi}=\ee^{-\xi a^{\dagger}}\ket{0}$,$\bra{\bar{\xi}}=\bra{0}\ee^{- a\bar{\xi}}$ are the related fermionic coherent states.
The time-reversal operation in this basis is \cite{ssr-17}
\begin{equation}
    \ket{\xi}\bra{\bar{\xi}} \longrightarrow \ket{\ii \bar{\xi}}\bra{\ii \xi}
    \label{eq:TF},
\end{equation}
and $\hat{\rho}_A^{R_2}$, the partial time-reversal of $\hat{\rho}_A$, is obtained by acting with \eqref{eq:TF} in \eqref{eq:gra} only in $A_2$. Although its spectrum is not real in general \cite{ssr-17,shapourian-19}, it  provides the fermionic (logarithmic) negativity as ${\cal E}= \log{\rm Tr} |\hat{\rho}_A^{R_1}|$.

In \cite{ssr-17}, the fermionic partial transpose has also been written in the occupation-number and in the Majorana fermion bases. These definitions are equivalent to a time-reversal operation up to a unitary trasformation: while this does not give any problems in the evaluation of the spectrum of $\hat{\rho}_A^{R_1}$, 
we should be careful if we are interested in the effect of partial transposition on the operators.

In this context, in \cite{mvdc-22} the negativity Hamiltonian has been introduced. The general field theoretical construction showed that it can be obtained from the knowledge of the entanglement Hamiltonian by performing a spatial inversion in the transposed intervals. To fix the ideas, let $[a_k, b_k]$ be the transposed interval. 
By applying the procedure introduced in \cite{cct-neg-1, cct-neg-2, cct-neg-3} for the partial transposition,
in the expression of the entanglement Hamiltonian the extremes $a_k$ and $b_k$ are exchanged and the function $w(x)$ is replaced by
\begin{equation}
    w^R(x) = \log \! \bigg[ \! - \frac{x-b_k}{x-a_k} \, \prod_{j\neq k} \frac{x-a_j}{x-b_j} \, \bigg].
\end{equation}
As a consequence, the local inverse temperature becomes $\beta^{R\text{ loc}}(x) = 1/w^{R}(x)'$ and the conjugate points $\tilde{x}_p^R$ will be the solutions of the equation $w^R(y) = w^R(x)$. We stress that the points $a_k$ and $b_k$ have to be exchanged also in the extremes of integration, which leads to an additional minus sign when the integration domain is the partially transposed interval.

It is immediate to extend the result of \cite{mvdc-22} to the presence of a boundary. In this case, we have to transpose both the interval $[a_k, b_k]$ and its reflection $[-b_k, -a_k]$ in the symmetric auxiliary geometry. This can again be implemented simply by exchanging the extremes $a_k$ and $b_k$ everywhere in the expression of the entanglement Hamiltonian with a boundary.

In \cite{mvdc-22}, the negativity Hamiltonian was written in the Majorana (real) fermion basis, $\mu$, where the fermionic partial transpose is implemented as $\mu(x) \to \ii \mu(x)$ for $x \in [a_k, b_k]$. Rewriting the
Dirac spinor $\psi$ in terms of two Majorana spinors as done in \eqref{eq:DiracMajorana}, we find out that the effect of the partial transposition is simply $\psi(x) \to \ii \psi(x)$, $\psi^{\dagger}(x) \to \ii \psi^{\dagger}(x)$ for $x \in [a_k, b_k]$. Let us stress that this transformation does not correspond to perform a time-reversal operation on the complex fermions, despite our starting point for the definition of the partial transposition in the coherent state basis.

For completeness, we report here the expression of the negativity Hamiltonian for two intervals $A=[a_1,b_1]\cup [a_2,b_2]$ on the real line obtained in \cite{mvdc-22} after the transposition of the second interval
\begin{equation}
\begin{aligned}\label{eq:NH-f}
    N_A&= \int_A dx\,\beta^{R\text{ loc }}(x)\,T_{00}(0,x) + \ii\left(\, \int_{a_1}^{b_1}-\int_{a_2}^{b_2} \,\right)dx\,\frac{\beta^{R\text{ loc }}(\tilde{x}^R)}{x-\tilde{x}^R}\; T^\text{bi-loc}_\text{diag}(x, \tilde{x}^R, 0),
\end{aligned}
\end{equation}
where 
\begin{equation}\label{eq:wbar}
\beta^{R\text{ loc }}(x)=\left(  \frac{1}{x-a_1}+\frac{1}{b_1-x}+\frac{1}{x-b_2}+ \frac{1}{a_2-x} \right)^{-1},
\end{equation}
and
\begin{equation}\label{eq:xbar}
    \tilde{x}^R=\frac{(b_1 a_2 - a_1 b_2) x + (b_1 + a_2) a_1  b_2 - (a_1 + b_2) b_1 a_2}{(b_1 - a_1 + a_2 - b_2) x + a_1 b_2 - b_1 a_2}.
\end{equation}
Since  $x$ and $\tilde{x}^R$ belong to different intervals, in the bi-local operator only one fermion receives a factor given by the imaginary unit under partial transposition. 
This implies that it is anti-hermitian. Moreover, because of the exchange of the extremes of integration, the bi-local term integrated over $[a_2, b_2]$ receives an additional minus sign.

In the following section, we show how to obtain the negativity Hamiltonian in the case of two intervals and one of them is adjacent to the boundary. 
For a higher number of intervals the procedure is completely analogous, albeit much more involved.

\subsection{Negativity Hamiltonian for two intervals with one adjacent to the boundary}
\label{sec:exampleNH}

Let us consider the subsystem $A=[0,a]\cup [b,c]$ on the half line. 
By transposing the second interval $[b,c]$, the function $w^R_\text{sym}(x)$ for the symmetric auxiliary geometry 
is obtained by exchanging $b$ and $c$ in the expression of $w_\text{sym}(x)$, finding 
\begin{equation}
    w^R_\text{sym}(x) = \log \left[\frac{(x+b)(x+a)(c-x)}{(x+c)(x-a)(x-b)}\right].
\end{equation}
Hence, the local effective inverse temperature reads
\begin{equation}\begin{split}\label{eq:betaR}
	&\beta^{R \text{ loc}}_\text{sym}(x) = \frac{1}{{w^R_\text{sym}}'(x)} = \left( \frac{2a}{x^2-a^2} + \frac{2b}{x^2-b^2} + \frac{2c}{c^2- x^2} \right)^{-1} .
\end{split}\end{equation}

The conjugate points are obtained as the solutions of the equation $w^R_\text{sym}(y)=w^R_\text{sym}(x)$.
Also for the negativity Hamiltonian we find a self-conjugate point $x^R_{S,2} \in [b,c]$, whose explicit expression is $x^R_{S,2} = \sqrt{ca+bc- ab}$. 
Similarly to what happened for the entanglement Hamiltonian, under the mapping $\tilde{x}_+^R$ the self-conjugate point $x^R_{S,2}$ is conjugated to the boundary.
This leads to a similar cross-over from the diagonal to the mixed operator in the bi-local term of the negativity Hamiltonian. 
The various intervals are mapped under conjugation as
\begin{equation}
    \begin{cases}
        \tilde{x}^R_+\!\left( [0, a] \right) = [b, x^R_{S,2}]\\
        \tilde{x}^R_+\!\left( [b, x^R_{S,2}] \right) = [0, a]\\
        \tilde{x}^R_+\!\left( [x^R_{S,2}, c] \right) = [-a, 0],
    \end{cases}\qquad \begin{cases}
        \tilde{x}^R_-\!\left( [0, a] \right) = [-x^R_{S,2}, -c]\\
        \tilde{x}^R_-\!\left( [b, c] \right) = [-c, -b]\,.
    \end{cases}
\end{equation}
Thus, in the bi-local operator, the terms corresponding to $\tilde{x}^R_+ ( [0, a] \cup [b, x^R_{S,2}] )$ contain fermions with the same chirality,
while the other ones provide terms that couple fermions with different chiralities, which also depend explicitly on the boundary condition. 

Moreover, under the partial transpose, the fermions receive an imaginary factor $\psi \to \ii \psi$ when they are evaluated in the transposed interval $[b,c]$. As a consequence, the bi-local operator that couples a point in $[0,a]$ with one in $[b,c]$ receives only one imaginary factor and is anti-hermitian.

Taking care of the sign of the conjugate points, the negativity Hamiltonian can be written more explicitly as follows
\begin{equation}\label{eq:nhTrip}\begin{split}
    N_A =& \int_{A} dx\, \beta^{R \text{ loc}}_\text{sym}(x)\, T_{00}(x, 0) 
    + \ii \, \bigg( \int_0^a - \int_b^{ x^R_{S,2}} \bigg) dx\, \frac{\beta^{R \text{ loc}}_\text{sym}(\tilde{x}^R_+)}{x-\tilde{x}^R_+}\, T^\text{bi-loc}_\text{diag}(x, \tilde{x}^R_+, 0) 
    \\
    &
     - \ii \int_{x^R_{S,2}}^c  dx\, \frac{\beta^{R \text{ loc}}_\text{sym}(\tilde{x}^R_+)}{x-\tilde{x}^R_+}\, T^\text{bi-loc}_\text{mix}(x, -\tilde{x}^R_+, 0; \alpha) 
    \\
    &
    + \left ( \ii \int_0^a +  \int_b^c \,\right ) dx\, \frac{\beta^{R \text{ loc}}_\text{sym}(\tilde{x}^R_-)}{x-\tilde{x}^R_-}\, T^\text{bi-loc}_\text{mix}(x, -\tilde{x}^R_-, 0; \alpha). 
\end{split}\end{equation}
We point out that the bi-local mixed term $T^\text{bi-loc}_\text{mix}(x, -\tilde{x}^R_-, 0; \alpha)$ 
contains $\psi(x)$ and $ \psi(-\tilde{x}_-^R)$ when integrated over $[b,c]$,
and they both take an imaginary unit factor under partial transposition because $x, -\tilde{x}_-^R \in [b,c]$. 
As a consequence, this operator is hermitian, differently from the same operator $T^\text{bi-loc}_\text{mix}(x, -\tilde{x}^R_-, 0; \alpha)$ integrated over $[0,a]$.

\section{Numerical lattice computations}\label{sec:lattice}

In order to check the validity of our predictions from CFT, in this section we compare them against exact numerical calculations. 
For free lattice fermions in the thermodynamic limit (i.e. for an infinite system), the exact two-point correlation matrix in the presence of a boundary is known \cite{Peschel2003,fc-11}. 
The relation \eqref{eq:peschel} between the reduced free fermion correlator $C_A$ and the entanglement Hamiltonian kernel $H_A$ provides the lattice entanglement Hamiltonian. 
However, recovering the continuum limit of the entanglement Hamiltonian from a discrete system turns out to be a non-trivial problem, as lengthy discussed in \cite{abch-16,ETP19,ETP22,EH-1,EH-2,EH-3,jt-21,defv-22}.
Indeed,  
the local inverse temperature has been obtained in some cases from the lattice by 
taking into account hoppings to distant neighbours and not only the nearest ones.

\subsection{Correlation matrix techniques}

Let us consider the tight-binding Hamiltonian
\begin{equation}
    \hat{H} = - \frac{1}{2} \sum_j \left( \hat{c}_j^\dagger \hat{c}_{j+1} + \hat{c}_{j+1}^\dagger \hat{c}_j \right),
\end{equation}
where the fermionic creation and annihilation operators $c_i, c_i^\dagger$ satisfy the canonical anticommutation relations
\begin{equation}
    \big\{ \hat{c}_i, \hat{c}_j^\dagger \big\}  = \delta_{ij}, 
    \qquad 
    \big\{ \hat{c}_i, \hat{c}_j \big\} = \big\{ \hat{c}_i^\dagger, \hat{c}_j^\dagger \big\} = 0.
\end{equation}

On the real line, the two-point correlation matrix is
\begin{equation}\label{eq:correlMatrix}
    \left ( C_A\right )_{i,j} = \frac{\sin\!\left (k_F (i-j) \right )}{\pi (i-j)},
\end{equation}
where $k_F$ is the Fermi momentum.
In the half chain, whose sites are labelled by $i \in \mathbb{N}$, 
we choose to impose open boundary conditions (OBC) on the first site (also known as Neumann boundary conditions).
The corresponding two-point correlation matrix has the following generic element \cite{Peschel2003, fc-11}
\begin{equation}\label{eq:correlMatrixOBC}
    \left ( C_A\right )_{i,j} = \frac{\sin\!\left (k_F (i-j) \right )}{\pi (i-j)} - \frac{\sin\!\left (k_F (i+j) \right )}{\pi (i+j)}.
\end{equation}
In our numerical computations we focus on the half-filled case, where $k_F = \frac{\pi}{2s}$, 
being $s$ the lattice spacing, that is set to $s=1$ in the numerical analysis.

In order to get the kernel $H_A$ in \eqref{eq:KA}, 
let us consider the discretised version of \eqref{eq:peschel}.
In this free fermionic model, the reduced density matrix can be written as \cite{Peschel2003, Peschel2004}
\begin{equation}\label{eq:latticeEH}
    \hat{\rho}_A 
    = \, \exp\!\big(\! -2\pi \hat{K}_A \big) 
    = \, \exp \! \bigg( \!\!-\sum_{i,j} \hat{c}_i^\dagger h_{i,j} \hat{c}_j \bigg).
\end{equation}
If $\sigma_k$ are the eigenvalues of the matrix $C_A$ and $\phi_k(j)$ its eigenvectors, 
from \eqref{eq:peschel} the finite-dimensional spectral representation of $h$ reads 
\begin{equation}\label{eq:spectralHA}
    h = V E V^\dagger,
\end{equation}
where the columns of $V$ are the eigenvectors $\phi_k$, while the matrix $E$ is diagonal and its elements are the eigenvalues $e_k$, related to the $\sigma_k$ as in \eqref{eq:lattPeschel}.
Because of the particle-hole symmetry, at half-filling both the correlation matrix of the infinite chain in \eqref{eq:correlMatrix} 
and of the half chain with OBC in \eqref{eq:correlMatrixOBC} 
have a checkerboard structure such that only matrix elements with $i-j$ odd are non-zero. 
This feature is inherited by $h$ \cite{ETP22} and such checkerboard structure greatly simplifies the continuum limit procedure, 
as we discuss later. 

We find it worth remarking that, in this numerical inspection, it is crucially important that the numerical values of $\sigma_k$ remain distinct from $0$ and $1$; hence the numerical analysis must be performed with high precisions.
We have used the python library mpmath \cite{mpmath}, keeping up to 300 digits for the lengths that we have considered.

The previous scheme can be adapted to the calculation of the negativity Hamiltonian \cite{mvdc-22}. 
In order to review this procedure, for simplicity we consider the subsystem $A = [a_1, b_1] \cup [a_2, b_2]$ made by the union of two disjoint blocks in the infinite chain.
The covariance matrix is defined as follows
\begin{equation}
    \Gamma_A 
    \,=\, \Id_A - 2 \,C_A \,=\, 
    \begin{pmatrix}
        \, \Gamma_{11}&    \Gamma_{12}\, \\
        \, \Gamma_{21}&    \Gamma_{22}\,
    \end{pmatrix},
\end{equation}
where $\Gamma_{11}$ and $\Gamma_{22}$ are the covariance matrices restricted to the blocks $[a_1, b_1] $ and $[a_2, b_2]$,
respectively, while $\Gamma_{12}$ and $\Gamma_{21}$ contain the cross correlations between them.
The partial time transposition of $[a_2, b_2]$  maps $\Gamma_A$ into a matrix 
whose generic element is the corresponding element of $\Gamma_A$ multiplied by an imaginary unit for each index belonging to $[a_2, b_2]$. The result is
\begin{equation}
\label{Gamma_R2_def}
    \Gamma_A^{R_2} 
    = \begin{pmatrix}
        \Gamma_{11}&  \!\!  -\ii \, \Gamma_{12}\\
        -\ii \, \Gamma_{21}&  \!\! -\Gamma_{22}
    \end{pmatrix}
    \equiv \,
    \Id_A - 2 \,C_A^{R_2}.
\end{equation}

The operator $\hat{\rho}_A^{R_2}$ obtained from $\hat{\rho}_A$ after a partial time transposition of the block $[a_2,b_2]$
reads
\begin{equation}\label{eq:latticeNH}
    \hat{\rho}_A^{R_2}
    = \, \exp\!\big(\! -2\pi \hat{N}_A \big) 
    = \, \exp \! \bigg( \!\!-\sum_{i,j} \hat{c}_i^\dagger \eta_{i,j} \hat{c}_j \bigg),
\end{equation}
where $\hat{N}_A$ is the negativity Hamiltonian,
whose kernel $\eta$ can be written in terms of the matrix $\Gamma_A^{R_2}$ in (\ref{Gamma_R2_def}) as follows
\begin{equation} \label{eq:peschelNeg}
    \eta 
    \,=\, \log \! \Big( \big[ \Id_A + \Gamma_A^{R_2} \big] \big[\Id_A - \Gamma_A^{R_2}\big]^{-1} \Big)    
        \,=\, \log \! \Big( \big[ C_A^{R_2}\big]^{-1} - \Id_A\Big).
\end{equation}
This matrix can again be obtained numerically by first solving the eigenvalue problem for $C_A^{R_2}$ and then employing its spectral representation. However, the difference with respect to \eqref{eq:spectralHA} is that $C_A^{R_2}$ and, consequently, $\eta$ are non-Hermitian. This implies that the matrix of the eigenvectors is non-unitary and \eqref{eq:spectralHA} needs to be modified as $\eta = V E V^{-1}$.

The connection between $\eta$ and $\Gamma_A^{R_2}$ was applied in \cite{mvdc-22} to evaluate the negativity Hamiltonian on the lattice and compare it with the field-theoretical prediction in \eqref{eq:NH-f}. In particular, in \cite{mvdc-22} the analytical local inverse temperature $\beta^{R \text{ loc}}(x)$ has been compared with the next-neighbour hopping term $\eta_{j, j+1}$ for the case of adjacent intervals, as we report in the inset of Fig.~\ref{fig:LocNegNoBound_diffInter}. 
While they are in good agreement with the local term of the field-theoretical prediction of \eqref{eq:NH-f} near the entangling points, 
a small deviation occurs as we move away from them. 
In the following we show that a perfect agreement also away from the entangling points is obtained 
by taking into account the higher hopping terms. 
This confirms the prediction in \eqref{eq:NH-f}.

\subsection{Continuum limit on the line}

The continuum limit of the entanglement Hamiltonian of a single block for a free fermion infinite chain in its ground state has been first obtained in \cite{ETP19}. 
Then, the procedure introduced in this work has been successfully adapted to recover the 
other entanglement Hamiltonians in free systems \cite{EH-1,EH-2,EH-3, jt-21}.

The core idea behind this limit is to express the model in terms of the low-energy fluctuations on top of the Fermi sea. 
This is done by introducing the continuous coordinate $x = i s$ and 
linearising the fluctuations of the lattice fermions $\hat{c}_i$ around the two Fermi points $\pm k_F = \pm \frac{\pi}{2s}$. 
In terms of the left- and right-moving fermions $\psi_L$ and $\psi_R$, we have that $\hat{c}_j$ reads \cite{ETP19}
\begin{equation}\label{eq:cj}
    \frac{\hat{c}_i}{\sqrt{s}} \sim e^{-\ii \frac{\pi}{2s} x} \psi_L(x) + e^{\ii \frac{\pi}{2s} x} \psi_R(x)\, .
\end{equation}
This expansion is the starting point of the limit studied in \cite{ETP22}, which is reviewed below.

Let us consider the subsystem made by the union of two disjoint segments $A=A_1 \cup A_2$ in the infinite chain. 
The corresponding entanglement Hamiltonian matrix is symmetric and has the following block structure
\begin{equation}\label{eq:block}
   h =  \begin{pmatrix}
        \, h^{(1,1)}&    h^{(1,2)} \, \\
        \, h^{(2,1)}&    h^{(2,2)} \,
    \end{pmatrix},
\end{equation}
where the diagonal blocks $h^{(1,1)}$ and $h^{(2,2)}$ describe hoppings within the first and second segment respectively, 
whereas the off-diagonal ones contain long-range hopping terms between the two segments.
This structure facilitates the continuum limit because the diagonal and the off-diagonal blocks in the entanglement Hamiltonian matrix (\ref{eq:block})
provide respectively the local and the bi-local terms of \eqref{eq:EHtot}, as discussed in \cite{ETP22}.

Let us first consider the diagonal blocks in (\ref{eq:block}),
whose elements are $h_{i, i+r}^{(\sigma, \sigma)}$ (with $\sigma \in \{1,2\}$).
Plugging \eqref{eq:cj} into the lattice entanglement Hamiltonian obtained from \eqref{eq:latticeEH}, one obtains
\begin{equation}\label{eq:locLimitDeriv1}
\begin{split}
    h_{i, i+r}^{(\sigma, \sigma)} \Big[ \hat{c}_i^\dagger \hat{c}_{i+r} + \hat{c}_{i+r}^\dagger \hat{c}_{i} \Big] 
    \sim 
    &\, s\, h_{i, i+r}^{(\sigma, \sigma)} \left [ e^{- \ii\frac{\pi}{2} r}  \psi_L^\dagger(x) \, \psi_L(x+rs) + e^{\ii\frac{\pi}{2} r} \,\psi_R^\dagger(x) \,\psi_R(x+rs)  \right .
    \\
    &
    \hspace{-.4cm}
    \left . +\, e^{\ii\frac{\pi}{2s} (2x + rs)} \psi_L^\dagger(x) \,\psi_R(x+rs) + e^{-\ii\frac{\pi}{2s} (2x + rs)} \psi_L^\dagger(x)\, \psi_R(x+rs) + \text{h.c.} \right ].
\end{split}
\end{equation}
The checkerboard structure of the matrix $h$ at half-filling 
tells us that the only the elements $h_{i,i+r}$ with $r$ odd are non-vanishing. 
We remark that the terms mixing different chiralities
are multiplied by a phase $\exp\!\left ( \pm \ii \frac{\pi}{2s} x\right )$ 
which is highly oscillating in the limit $s \to 0$, for fixed $x$. 
Since $\psi_L, \psi_R$ are smooth functions, such mixed terms cancel out in the continuum limit, 
leading to the decoupling of left- and right-movers expected for conformally invariant models on the line. 
Thus, in \eqref{eq:locLimitDeriv1} we can drop these terms
and expand both $\psi_L, \psi_R$ and the matrix element $h_{i, i+r}$ in powers of the lattice spacing $s$, finding \cite{ETP19, ETP22}
\begin{multline}
\label{eq:locLimitDeriv2}
    h_{i, i+r}^{(\sigma, \sigma)} \left [ \hat{c}_i^\dagger \hat{c}_{i+r} + \hat{c}_{i+r}^\dagger \hat{c}_{i} \right ] \approx 
    \\
   \approx \, s\, h_{i-\frac{r-1}{2}, i+\frac{r+1}{2}}^{(\sigma, \sigma)} 
   \left [\,- \ii\, \sin\!\left ( \frac{\pi}{2} r \right ) r\, s \left ( \psi_L^\dagger(x) \,\partial_x \psi_L(x) - \psi_R^\dagger(x) \,\partial_x \psi_R(x) \right )  + \text{h.c.} \right ],
   \end{multline}
   where in the r.h.s. we recognise the energy density $T_{00}$ of the Dirac fermion reported in \eqref{eq:energydensity} at $t=0$.
Plugging first \eqref{eq:locLimitDeriv2} in the expression of the lattice entanglement Hamiltonian in \eqref{eq:latticeEH},
and then sending $s \to dx$, $i s \to x$ and the sum over the index $i$ to a spatial integral, one obtains
\begin{equation}\label{eq:limitLocalEH}
    \frac{1}{2\pi} \sum_{i, j} h_{i,j}^{(\sigma, \sigma)} \hat{c}_i^\dagger \hat{c}_j \sim \frac{1}{2\pi} \int dx\, \mathcal{S}^\text{loc}(x)\, T_{00}(x),
\end{equation}
where \cite{ETP19, ETP22}
\begin{equation}\label{eq:limitLocalAntidiag}
    \mathcal{S}^{\mathrm{loc}}(x)\equiv
     - \,2 s \sum_{\substack{r\text{ odd}\\ r\geq 1}} r\, (-1)^{\frac{r-1}{2}} h_{i-\frac{r-1}{2}, i+\frac{r+1}{2}}^{(\sigma, \sigma)}\,.
\end{equation}

In \cite{ETP22} it has been checked numerically that
 $\mathcal{S}^{\mathrm{loc}}$ in \eqref{eq:limitLocalAntidiag} converges to $2\pi \beta^\text{loc}(x)$ in \eqref{eq:betalocgen} in the continuum limit. 
We remark that, according to \eqref{eq:limitLocalAntidiag},
this continuum limit requires to sum over higher hoppings
and not to consider only the next-neighbour element $h_{i, i+1}$, 
as one would expect from a naive discretisation of the stress-energy tensor in \eqref{eq:EHtot}.

Recently \cite{ETP22}, the previous considerations have been extended 
in order to obtain the bi-local terms of the entanglement Hamiltonian \eqref{eq:EHtot} through a continuum limit. 
Let us recall that, when $A$ is the union of disjoint intervals, 
the bi-local term in \eqref{eq:EHtot} (see also \eqref{eq:qlocdiag}) couples fermions evaluated in two different positions
given by $x$ and a certain conjugate point $\tilde{x}_p(x)$ satisfying \eqref{eq:w} and belonging to a different interval.
Thus, e.g. for $A = A_1 \cup A_2$ and $x\in A_1$ we have that $\tilde{x}_1 \in A_2$ and therefore the off-diagonal block $h_{i,j}^{(1,2)}$
in the lattice entanglement Hamiltonian matrix (\ref{eq:block}) must be considered. 
In analogy with the field-theoretical result, the non-zero matrix elements of this off-diagonal block turns out to be localised around the curve defined by $js =\tilde{x}_p(is)$
\cite{abch-16, ETP22}.
However, despite the fact that the matrix elements decay as we move away from this curve, they remain non-zero even far from it, 
similarly to what happens for the diagonal blocks $h^{(\sigma, \sigma)}$ previously discussed. 
This necessary leads to take into account all the elements of the off-diagonal block in order to reproduce the correct result in the continuum limit \cite{ETP22}.

Plugging the expression for $\hat{c}^{\dagger}_j$ and $\hat{c}_j$ given by  \eqref{eq:cj} 
into the entanglement Hamiltonian obtained from \eqref{eq:latticeEH},
the term provided by the off-diagonal block $h^{(1,2)}$ reads \cite{ETP22}
\begin{equation}\begin{split}\label{eq:expansionBilocal}
    \hat{c}_i^\dagger h_{i,j}^{(1,2)} \hat{c}_j\,
    \sim&\; s\, h_{i,j}^{(1,2)}\, \Big [e^{\ii \frac{\pi}{2} (i-j) } \psi_L^\dagger(x) \, \psi_L(y) + e^{\ii  \frac{\pi}{2} (j-i) } \psi_R^\dagger(x) \, \psi_R(y)  
    \\
    & \hspace{1.4cm} + e^{\ii \frac{\pi}{2} (i+j) } \psi_L^\dagger(x)\, \psi_R(y) + e^{- \ii \frac{\pi}{2} (i+j) } \psi_R^\dagger(x) \, \psi_L(y) \Big ]  
    \\
    \rule{0pt}{.8cm}
    =&\; \, \ii\, s\,   \sin\!\left ( \frac{\pi}{2} (j-i) \right ) h_{i,j}^{(1,2)} \left [ \psi_R^\dagger(x) \, \psi_R(y) - \psi_L^\dagger(x) \, \psi_L(y) \right ] 
    \\
    \rule{0pt}{.6cm}
    & \hspace{.2cm}  -\ii\, s\,   \sin\!\left ( \frac{\pi}{2} (i+j) \right ) h_{i,j}^{(1,2)}  \left [ \psi_R^\dagger(x) \, \psi_L(y) - \psi_L^\dagger(x) \, \psi_R(y) \right ],
\end{split}\end{equation}
where in the second equality the fact that the elements $h_{i,j}^{(1,2)}=0$ when $i \pm j$ is even has been employed. 
The terms multiplied by $\sin(\frac{\pi}{2} (i+j))$ in (\ref{eq:expansionBilocal}) are highly oscillating, hence they can be dropped in the continuum limit.
Expanding the fermions $\psi_L, \psi_R$ in powers of the lattice spacing around the conjugate point $\tilde{x}_p(x = is)$ and keeping only the zeroth order, one obtains
\begin{equation}\label{eq:expansionBilocal2}
    \hat{c}_i^\dagger h^{(1,2)}_{i,j} \hat{c}_j 
    \approx \ii\, s\,   \sin\!\left ( \frac{\pi}{2} (j-i) \right ) h^{(1,2)}_{i,j} \left [ \psi_R^\dagger(x) \psi_R(\tilde{x}_p) - \psi_L^\dagger(x) \psi_L(\tilde{x}_p)\right ],
\end{equation}
where in the r.h.s. we recognise the bi-local operator $T^\text{bi-loc}_\text{diag}(x,\tilde{x}_p,t=0)$ defined in \eqref{eq:qlocdiag}, which does not mix fields with different chiralities. 
Thus, in the continuum limit \eqref{eq:expansionBilocal2} provides the following term 
in the lattice entanglement Hamiltonian (see \eqref{eq:latticeEH})
\begin{equation}\label{eq:limitBilocDiagEH}
    \frac{1}{2\pi} \sum_{i,j} \hat{c}_i^\dagger h_{i,j}^{(1,2)} \hat{c}_j
    \sim \frac{1}{2\pi} \int_{a_1}^{b_1} dx\, \mathcal{S}^\text{diag}(x)\, T^\text{bi-loc}_\text{diag}(x, \tilde{x}_p, t=0),\quad \text{with } \tilde{x}_p \in [a_2, b_2],
\end{equation}
where, taking advantage of the checkerboard structure of $h_{ij}$, we have introduced \begin{equation}\label{eq:limitBilocDiag}
    \mathcal{S}^\text{diag}(x) \equiv \sum_{j} (-1)^{(j-i-1)/2} h_{i,j}^{(1,2)} .
\end{equation}
A similar result is obtained for the off-diagonal block matrix $h_{i,j}^{(2,1)}$ in (\ref{eq:block}).

In Ref.~\cite{ETP22} it has been checked numerically that the continuum limit of (\ref{eq:limitBilocDiag}) and of the corresponding expression involving $h_{i,j}^{(2,1)}$
provide the weight function \eqref{eq:bilocalweight}  of the bi-local term of the entanglement Hamiltonian \eqref{eq:EHtot} in the case of two intervals.
Thus, like for the local term, the whole block matrices $h_{i,j}^{(1,2)}$ and $h_{i,j}^{(2,1)}$ must be used to get the field theory prediction, 
and not only the matrix elements around $\tilde{x}_p(x)$, as one could naively expect.

\subsubsection{Negativity Hamiltonian}\label{susec:NHinfinite}

\begin{figure}[t!]
    \centering
    {\includegraphics[width=.49\textwidth]{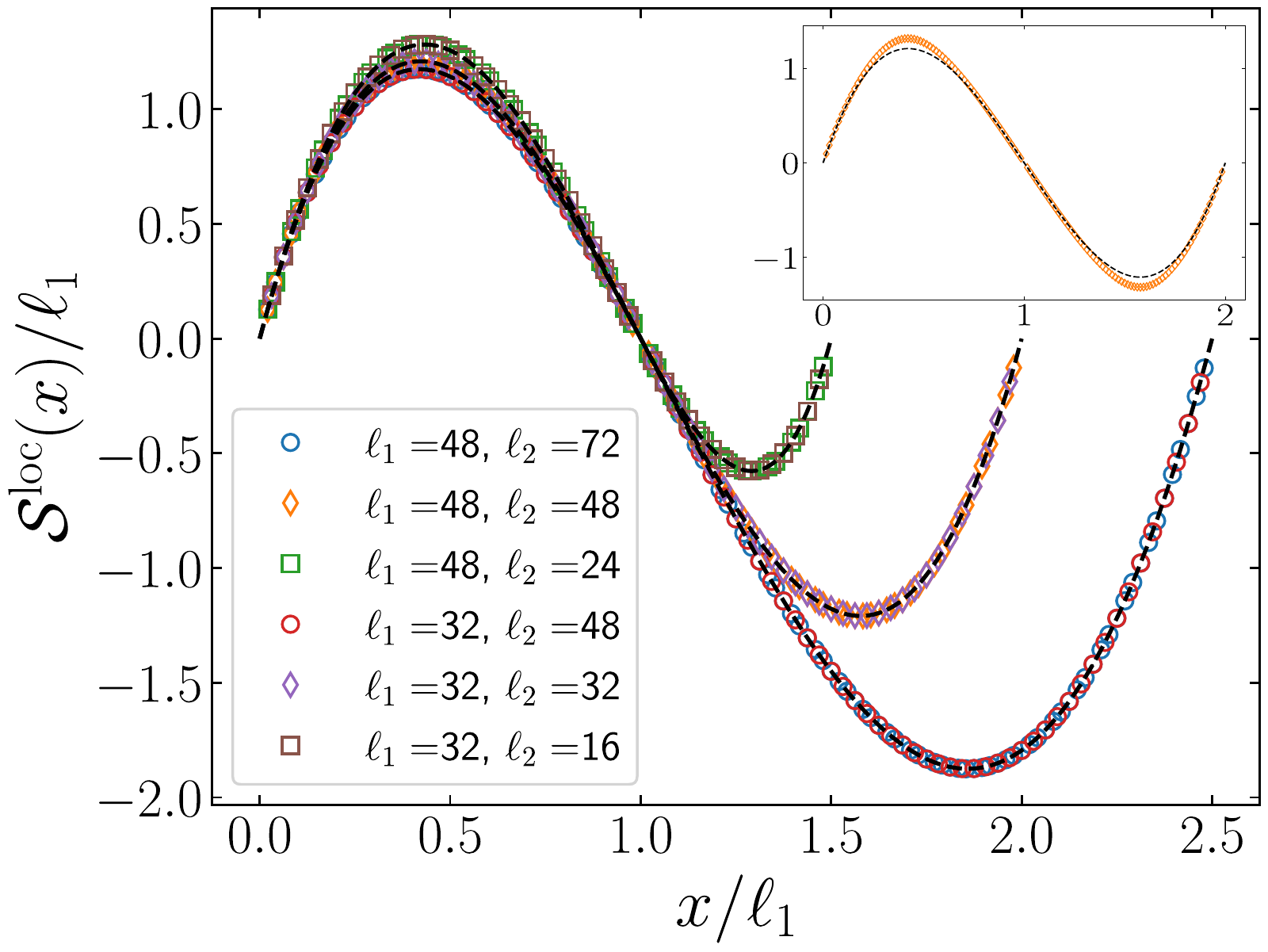}}
    \subfigure
    {\includegraphics[width=.49\textwidth]{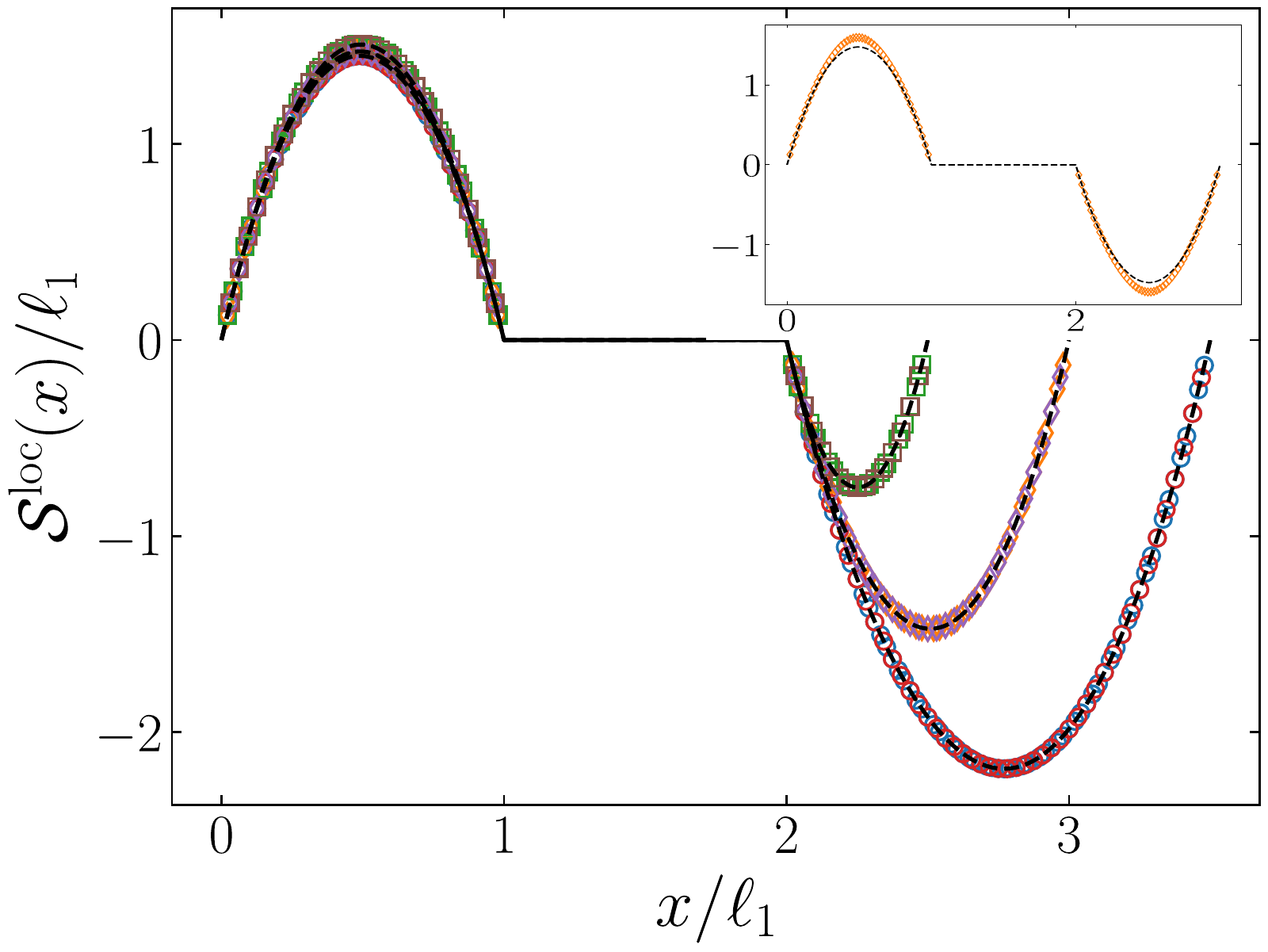}}
    \newline
    \subfigure
    {\includegraphics[width=.49\textwidth]{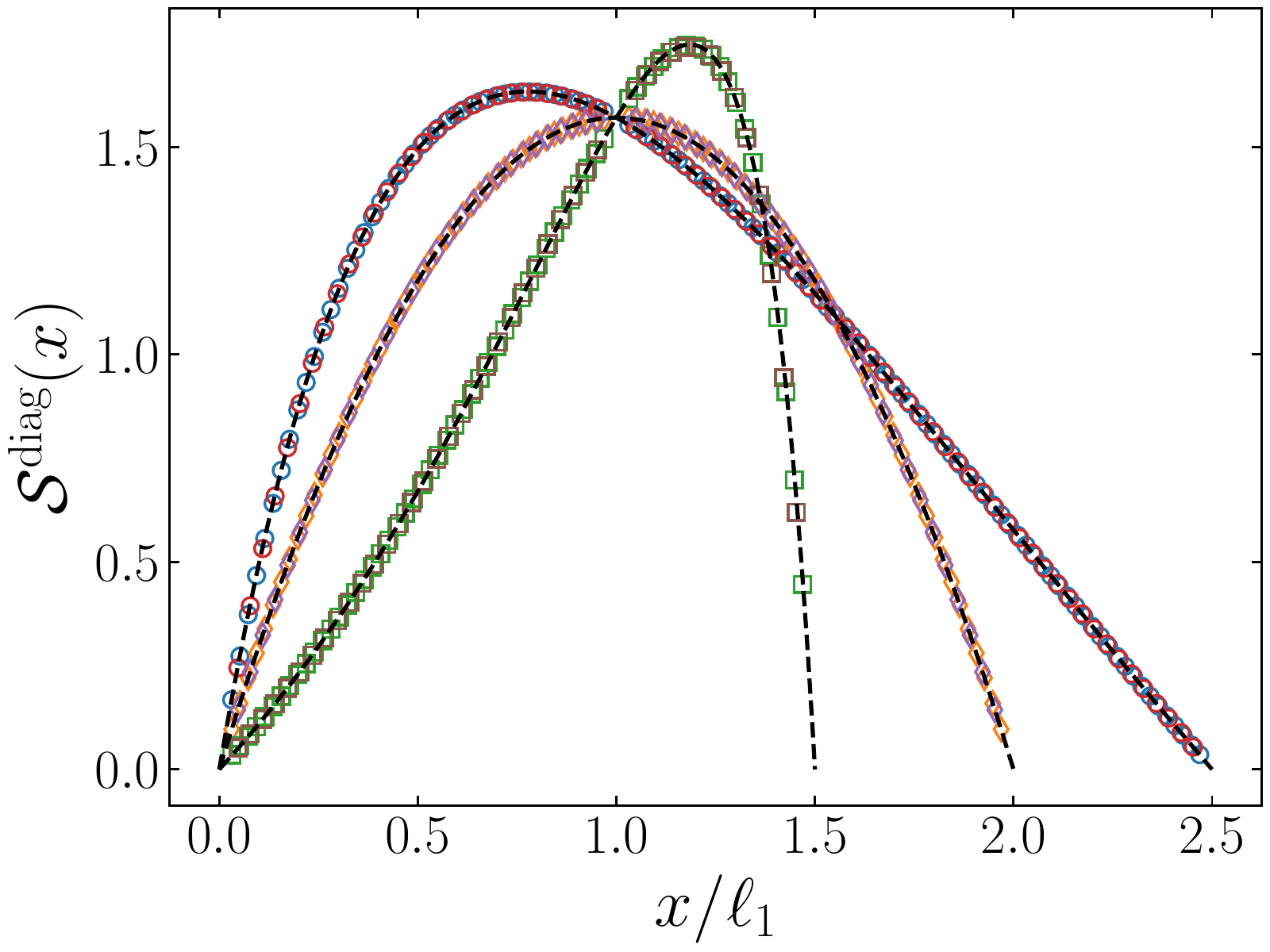}}
    \subfigure
    {\includegraphics[width=.49\textwidth]{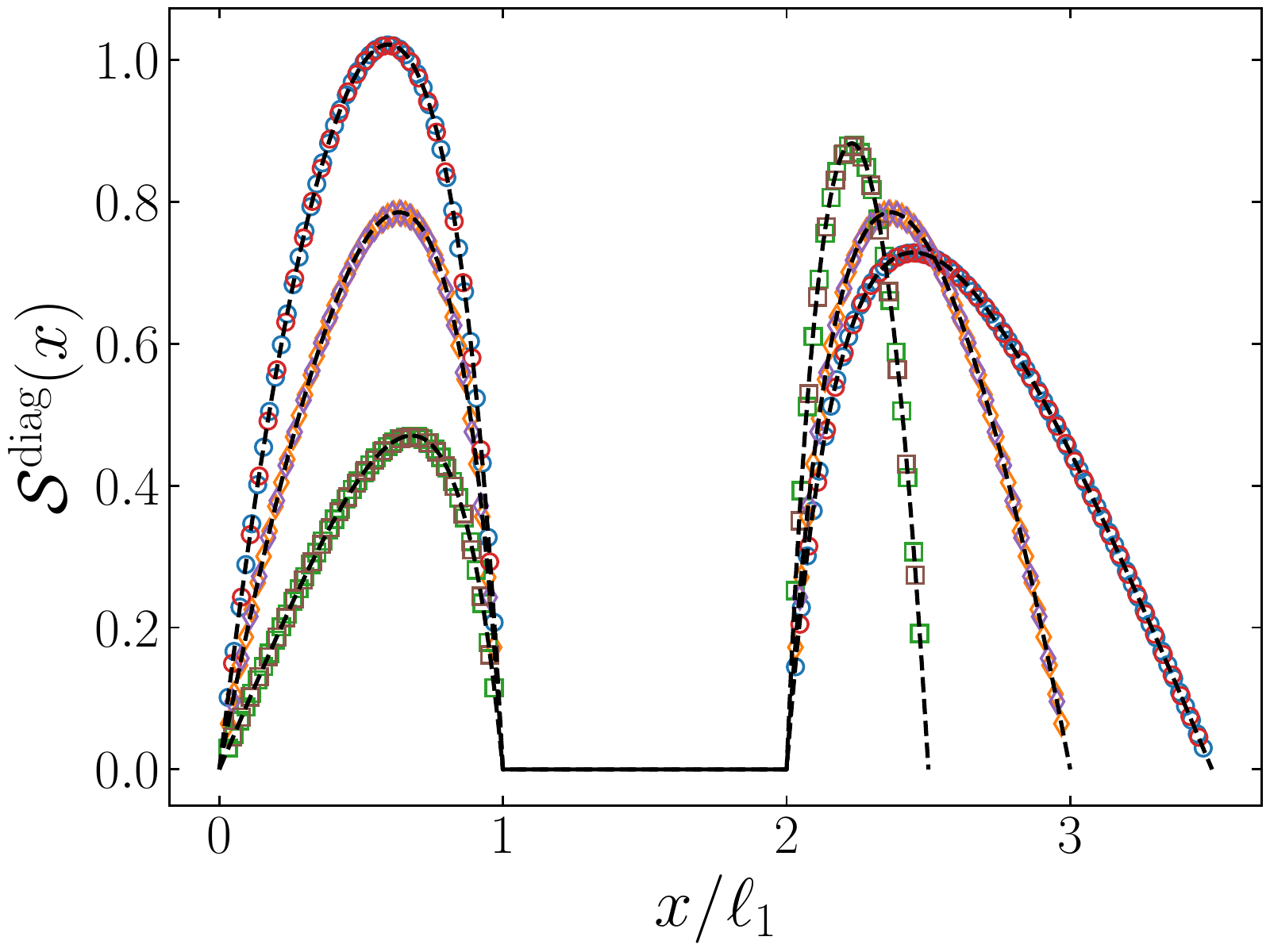}}
    \caption{
    Benchmark of the analytical prediction for the negativity Hamiltonian of adjacent (left panels) and disjoint (right panels) blocks in the infinite chain for a Dirac fermion. 
    We consider a subsystem $A= A_1 \cup A_2$, with $A_1=[1,\ell_1] \cup A_2=[\ell_1+d,\ell_1+\ell_2]$, where $d=1,\ell_1$ on the left and right panels, respectively, for different length ratios $\ell_2/\ell_1 = 1.5, 1, 0.5$. 
    In the top panels, the symbols are obtained from \eqref{eq:limitLocalAntidiag} while the dashed lines correspond to \eqref{eq:wbar}, rescaled by $\ell_1$ in order to show the collapse for different sizes. 
    The insets show that considering only the nearest neighbours does not provide a perfect agreement away from the entangling points. 
    In the bottom panels, the symbols are obtained from \eqref{eq:limitBilocDiag} while the dashed line corresponds to 
    the weight function in the bi-local term in \eqref{eq:NH-f}.
    }
    \label{fig:LocNegNoBound_diffInter}
\end{figure}

The formulae \eqref{eq:limitLocalAntidiag} and \eqref{eq:limitBilocDiag} have been obtained in \cite{ETP19, ETP22} 
to find the continuum limit of the entanglement Hamiltonian for a multipartite geometry. 
We remark that, 
in the derivations of \eqref{eq:locLimitDeriv1}, \eqref{eq:locLimitDeriv2} for the local part and of \eqref{eq:expansionBilocal}, \eqref{eq:expansionBilocal2} for the bi-local one, 
only the expansion \eqref{eq:cj} of the lattice fermion $\hat{c}_i$ in terms of the low energy fluctuations $\psi_L, \psi_R$ has been used. 
Such expansion is valid in general, not only for the entanglement Hamiltonian;
hence the steps of the previous section can be repeated for the lattice negativity Hamiltonian \eqref{eq:latticeNH}, 
by replacing each block of $h_{i,j}$ with the corresponding block of the negativity kernel $\eta_{i,j}$, which also inherits a checkerboard structure. 
This allows us to find that the continuum limit of the lattice negativity Hamiltonian  has 
a local term whose weight function can be read from \eqref{eq:limitLocalAntidiag}, 
while the bi-local terms take different signs and imaginary factors in different intervals.
In the special case of two intervals, this can be see from \eqref{eq:NH-f},
where the bi-local term is the imaginary part of $N_A$ and the integral over $[a_2, b_2]$ has an additional minus sign due to the partial transposition. 
This tells us that, in order to compare the continuum limit of the lattice negativity Hamiltonian with the field-theoretical prediction of \eqref{eq:NH-f}, 
\eqref{eq:limitBilocDiag} must be modified as follows
\begin{equation}\label{eq:limitBilocDiagNeg}
    \mathcal{S}^\text{diag}(x) = \begin{cases}
       \;  - \ii \sum_{j} (-1)^{(j-i-1)/2} \eta_{i,j}^{(1,2)}   \;\;\; &x \in [a_1, b_1]
        \\
        \rule{0pt}{.8cm}
       \;  \ii \sum_{j} (-1)^{(j-i-1)/2} \eta_{i,j}^{(2,1)}    &x \in [a_2, b_2]\,.
    \end{cases}
\end{equation}
Now we can study the continuum limit of \eqref{eq:limitLocalAntidiag} and \eqref{eq:limitBilocDiagNeg} 
to check the field theory predictions for the negativity Hamiltonian, 
which is  reported in \eqref{eq:NH-f}, \eqref{eq:wbar} and \eqref{eq:xbar} 
for two disjoint intervals of arbitrary length, after a partial transposition of the second one.

In Fig.~\ref{fig:LocNegNoBound_diffInter} we consider three different length ratios $\ell_2/\ell_1 = 0.5, 1, 1.5$ for two adjacent intervals (left panels) or two disjoint intervals separated by $\ell_1$ sites.
As for the continuum limit of the diagonal blocks \eqref{eq:limitLocalAntidiag}, 
in the top panels we find that the sum $\mathcal{S}^\text{loc}$ over the higher hoppings is in perfect agreement with the field-theoretical local effective inverse temperature in \eqref{eq:wbar}, even away from the entangling points.
We recall that in \cite{mvdc-22} only the nearest neighbour negativity Hamiltonian has been considered; 
hence this accurate test of the continuum limit,
which involves also long-range hoppings, appears here for the first time.

As for the non-local term of the negativity Hamiltonian, in \cite{mvdc-22} the numerical calculation of the bi-local weight function was limited 
to the simple case in which the two intervals have the same length and the main contribution to the bi-local term of \eqref{eq:NH-f} comes from the antidiagonal elements of $\eta$. 
However, for arbitrary lengths of the two intervals, it is even more difficult to select the matrix elements corresponding to the bi-local term and, as a consequence, 
a proper continuum limit is necessary to recover the field theory results.
In the bottom left (bottom right) panel of Fig.~\ref{fig:LocNegNoBound_diffInter}, 
we compare $\mathcal{S}^\text{diag}$ in \eqref{eq:limitBilocDiagNeg} for adjacent (non-adjacent) intervals and different ratios of their lengths with the field-theoretical weight function $\beta^{R\text{ loc } }(\tilde{x}^R)/(x-\tilde{x}^R)$ occurring in the bi-local term of the negativity Hamiltonian in \eqref{eq:NH-f}.
In all cases perfect agreement is obtained between the field theory expression \eqref{eq:NH-f} and the numerical results.

\subsection{Continuum limit in the presence of a boundary}

In the following we discuss how the continuum limit procedure described above is modified in the presence of a boundary. 
In Sec.~\ref{sec-one-int-bdy} we review the case of the entanglement Hamiltonian of one interval considered in \cite{ETP22}.
Then we apply this continuum limit procedure to the subsystem made by two intervals with the first one adjacent to the boundary 
(see Sec.~\ref{sec:example} and Sec.~\ref{sec:exampleNH}), 
in order to recover numerically the weight functions occurring in 
the entanglement Hamiltonian (Sec.~\ref{sec-two-int-bdy-eh})
and in the negativity Hamiltonian associated to the partial transposition of the second interval (Sec.~\ref{sec-two-int-bdy-en}).

\subsubsection{Entanglement Hamiltonian: single interval}
\label{sec-one-int-bdy}

In order to describe the ingredients needed to benchmark the theoretical predictions of this manuscript, 
let us briefly review the result of \cite{ETP22},
where the proper continuum limit of the lattice entanglement Hamiltonian for one single interval $A=[b,c]$ in the presence of the boundary
has been studied. 
In this case the lattice entanglement Hamiltonian $h$ consists of one single block 
and the corresponding field theoretical prediction in \eqref{eq:limit1}  is the sum of two terms: 
a local one  proportional to the stress-energy tensor \eqref{eq:energydensity}
and a bi-local one proportional to the operator $T^\text{bi-loc}_\text{mix}$ in \eqref{eq:mixBilocOp} 
that mixes fields with different chiralities.

As for  the local part of the entanglement Hamiltonian, in \cite{ETP22} it has been found that
the combination of the matrix elements of $h$ to consider is
\begin{equation}\label{eq:limitLocalRows}
    \mathcal{S}^\text{loc}(x) \equiv - 2 s \sum_r r\,(-1)^{(r-1)/2} h_{i,i+r}\, ,
\end{equation}
which differs from \eqref{eq:limitLocalAntidiag} only up to higher order terms in the lattice spacing
and may introduce slight deviations, as discussed in \cite{EH-2} for the harmonic chain.

For the continuum limit of the bi-local term, as discussed in \cite{ETP22}, the correlator in \eqref{eq:correlMatrixOBC} for OBC corresponds to the vector phase condition \eqref{eq:vectorCond} with scattering phase $\alpha = \pi$, that is
\begin{equation}
    \psi_R(x=0) = - \, \psi_L(x=0).
\end{equation}
Therefore, using in \eqref{eq:limitBilocDiagEH} the bi-local operator in the vector phase of \eqref{eq:limit1} and replacing this result in the expression of the lattice entanglement Hamiltonian \eqref{eq:latticeEH}, we obtain in the continuum limit
\begin{equation}\label{eq:limitBilocEHBound1int}\begin{split}
    &\frac{1}{2\pi} \sum_{i,j} \hat{c}_i^\dagger h_{i,j} \hat{c}_j
    \sim \frac{1}{2\pi} \int_{A} dx\, \mathcal{S}^\text{loc}(x)\, T_{00}(x)\,
     \\ 
    &\hspace{2.8cm}
    + \frac{1}{2\pi} \int_{A} dx\, \mathcal{S}^\text{mix}(x)\, T^\text{bi-loc}_\text{mix, vec}(x, -\tilde{x}^{}, 0; \alpha = \pi) ,\quad \text{with } -\tilde{x}^{}=\frac{bc}{x}, 
\end{split}\end{equation}
where $\mathcal{S}^\text{loc}$ is given by \eqref{eq:limitLocalRows} and we have introduced the sum over the columns of the matrix $h$
\begin{equation}\label{eq:limitBilocMix}
    \mathcal{S}^\text{mix}(x) \equiv \sum_j (-1)^{(i+j-1)/2} h_{i,j}.
\end{equation}
For different boundary conditions, this expression takes a different form, as discussed in \cite{ETP22}.

An important complication due to the presence of a boundary is that the non-diagonal bi-local operator and the local one are superimposed, making it difficult to distinguish their different weight functions. 
A solution to this problem was proposed in Ref.~\cite{ETP22} by observing that, from
\eqref{eq:limitLocalAntidiag} and \eqref{eq:limitBilocMix}, the matrix elements $h_{i,j}$ contributing to the two weight functions have distinct oscillating phases that cannot be compensated at the same time.
This allows us to isolate, for example, the local term by averaging the $i$-th term with its nearest neighbours as 
\begin{equation}\label{eq:avgTrick}
    \widetilde{\mathcal{S}}^{\mathrm{loc}}(i) = \frac{1}{4}\,\mathcal{S}^{\mathrm{loc}}(i-1) + \frac{1}{2}\,\mathcal{S}^{\mathrm{loc}}(i)  + \frac{1}{4}\,\mathcal{S}^{\mathrm{loc}}(i+1),
\end{equation}
because the unwanted contribution is an alternating function eliminated through such average.
A similar strategy can be applied for $\mathcal{S}^\text{mix}$. 
In Ref.~\cite{ETP22}, by using the average in \eqref{eq:avgTrick} and the one corresponding to $\mathcal{S}^\text{mix}$, 
it was checked numerically that the weight functions 
in the field-theoretical entanglement Hamiltonian \eqref{eq:limit1} are obtained in the continuum limit.

\subsubsection{Entanglement Hamiltonian: two intervals}
\label{sec-two-int-bdy-eh}

\begin{figure}[t!]
    \centering
    {\includegraphics[width=1\textwidth]{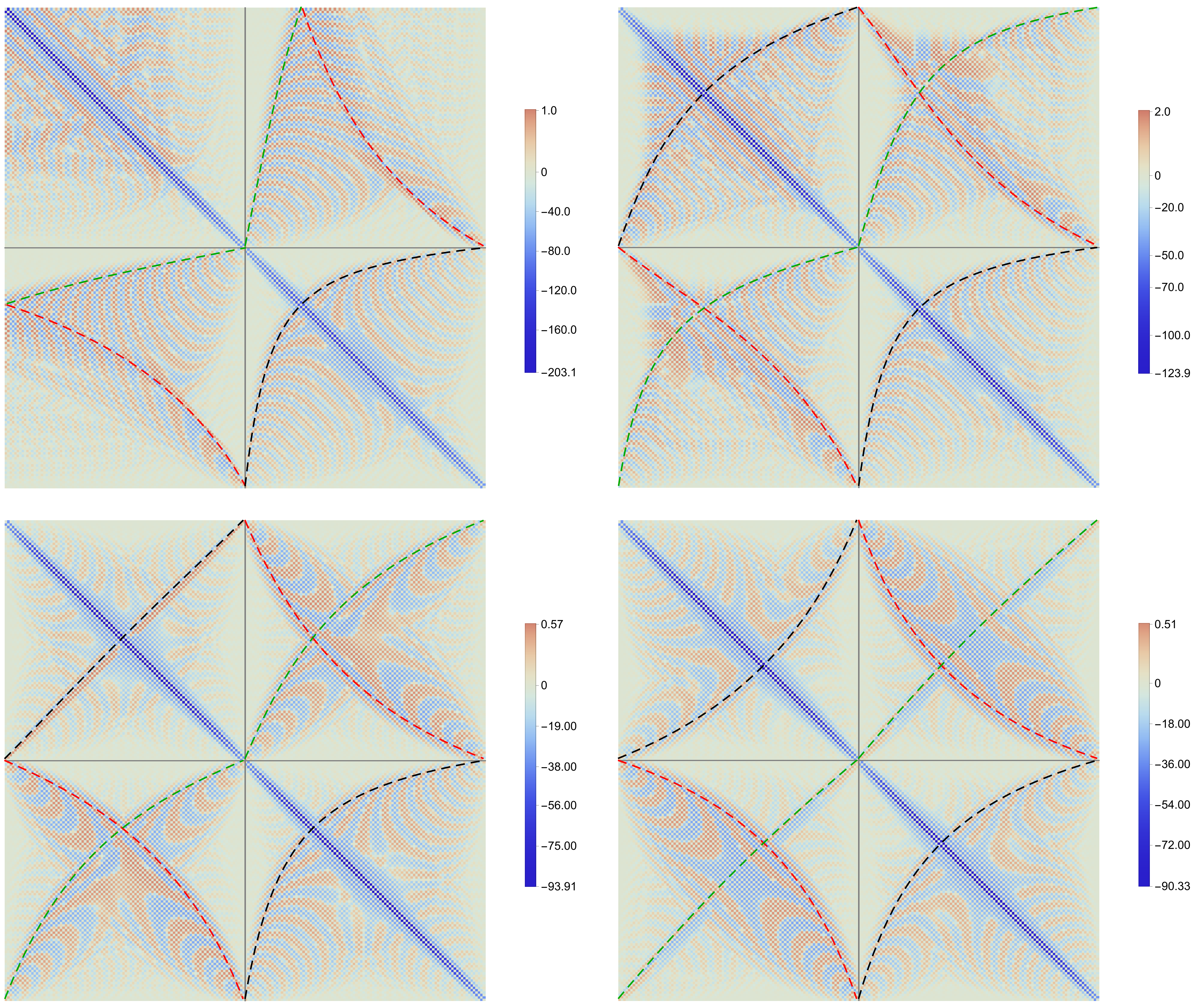}}
    \caption{
        Matrix elements of the entanglement Hamiltonian matrix $h$ for two disjoint intervals in the semi-infinite chain with OBC,
        made by $\ell_1=\ell_2=100$ consecutive sites and separated by $100$ consecutive sites.
        The distance between the boundary and the first interval is given by $0$ (top left), $10$ (top right), $100$ (bottom left) and $1000$ (bottom right) consecutive sites.
        The dashed lines correspond to the conjugate points (see Fig.~\ref{fig:aux_geometry}).}
    \label{fig:MP-plot}
\end{figure}

Consider the subsystem $A$ given by the union of two disjoint intervals in the semi-infinite chain.
The corresponding entanglement Hamiltonian matrix $h$
which has the block structure of \eqref{eq:block},
is visualised in Fig.~\ref{fig:MP-plot}
for the special case where the lengths of the intervals and their distance take the same value.
The dashed lines display the position of the conjugate points.
For the sake of simplicity, we focus on the case where the first interval is adjacent to the boundary,
namely $A = [0,a] \cup [b,c]$ (top left panel in Fig.~\ref{fig:MP-plot}).
The corresponding field-theoretical results for the entanglement Hamiltonian are 
reported in Sec.~\ref{sec:example}.

\begin{figure}[t!]
    \centering
    \includegraphics[width=.7\textwidth]{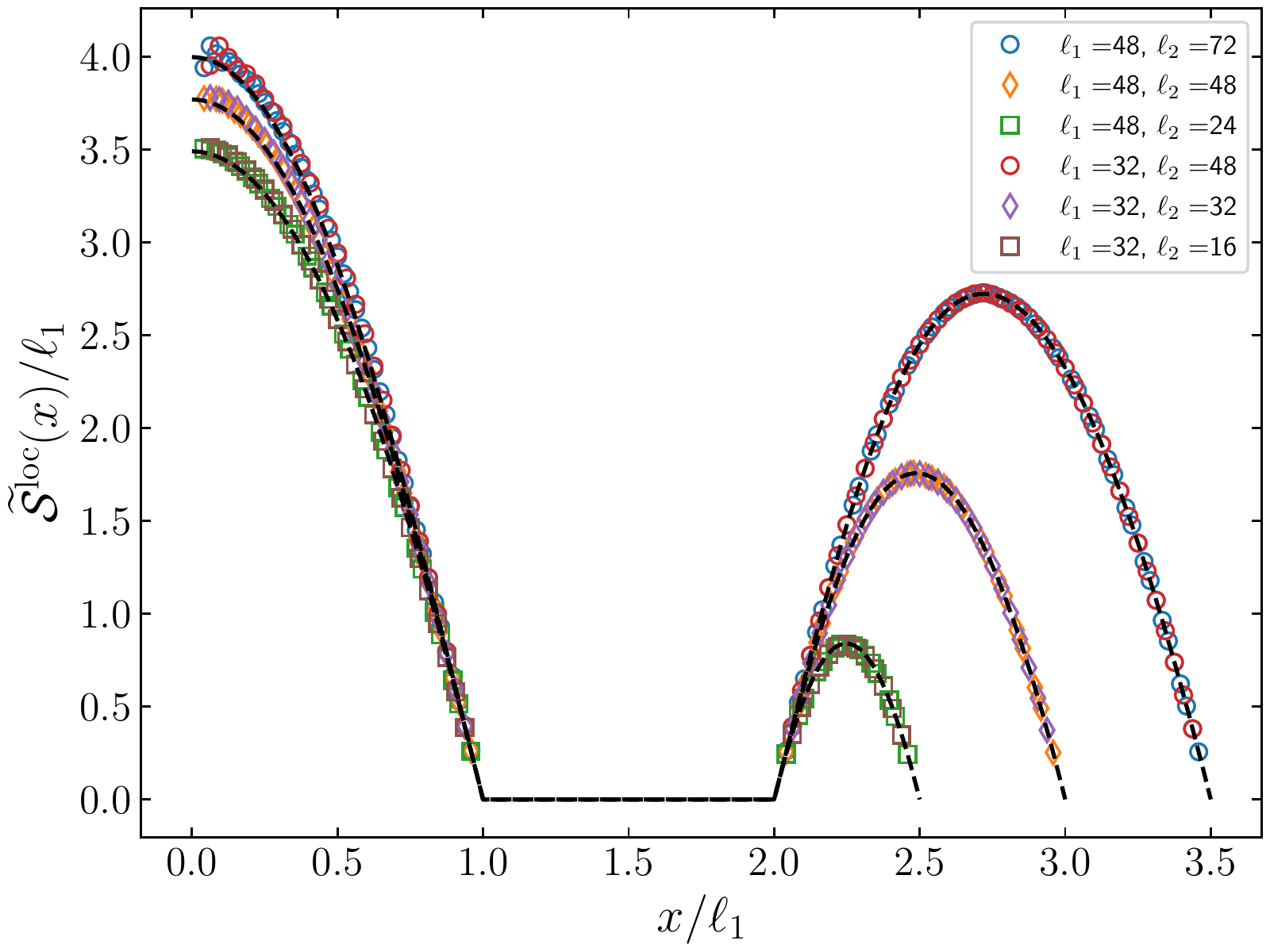}
    \caption{Local effective temperature of the entanglement Hamiltonian for the union of two disjoint intervals
    $A=[1,\ell_1] \cup [2\ell_1+1,2\ell_1+\ell_2]$ in the semi-infinite chain with OBC,
    for different values of the ratio $\ell_2/\ell_1 = 0.5, 1, 1.5$.
    The dashed line corresponds to the CFT prediction \eqref{eq:betaLocTwoBound}.
    The numerical data (symbols) are obtained by using \eqref{eq:limitLocalRows} for the first interval and \eqref{eq:limitLocalAntidiag} for the second one by averaging over the neighbouring sites according to \eqref{eq:avgTrick}.
    }
    \label{fig:LocEnt_diffInter}
\end{figure}

In Fig.~\ref{fig:LocEnt_diffInter} we show numerical results about the local term, 
which corresponds to the blocks $h^{(1,1)}$ and $h^{(2,2)}$ (see \eqref{eq:block}). 
The continuum limit is performed by employing 
\eqref{eq:limitLocalRows} for the interval $[0,a]$ and \eqref{eq:limitLocalAntidiag} for $[b,c]$,
both combined with the average over the neighbours as in \eqref{eq:avgTrick}.
Different ratios $\ell_2/\ell_1 = 0.5, 1, 1.5$ are considered, where $\ell_1=a, \ell_2=c-b$ are the lengths of the two intervals,
keeping the distance between them $b - a$ fixed and equal to $\ell_1$.
The data are rescaled by $\ell_1$ in order to show a collapse for different sizes.
Perfect agreement with the field-theoretical local effective inverse temperature of 
\eqref{eq:betaLocTwoBound} is obtained.

\begin{figure}[t!]
\vspace{.8cm}
    \centering
    \includegraphics[width=1.0\textwidth]{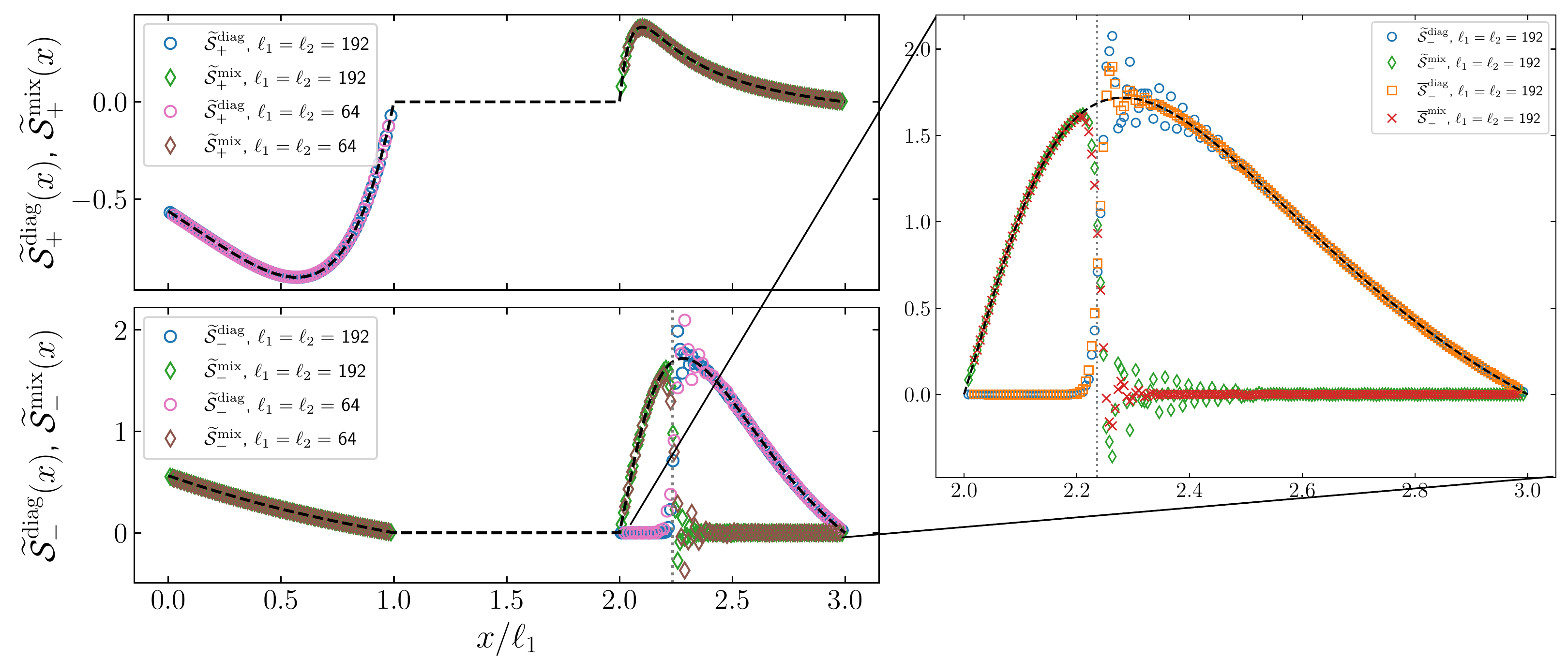}
    \caption{
    Scaling of the bi-local weights of the entanglement Hamiltonian as a function of $x/\ell_1$ for the geometry $A=[1,\ell_1] \cup [2\ell_1+1,3\ell_1]$ in the presence of a boundary at $x=0$, with $\ell_1=64, 192$.
    The top (bottom) panel shows the bi-local weight function relative to $\tilde{x}_+$ ($\tilde{x}_-$), while the different colours distinguish between the non-diagonal and diagonal operators according to the legend. The green/brown (blue/pink) symbols have been obtained by 
    applying \eqref{eq:avgTrick} 
    to \eqref{eq:limitBilocMix} (\eqref{eq:limitBilocDiag}).
    The dotted grey line indicates the self-conjugate point $x_{S,2}$ in \eqref{eq:selfpoint}, while the dashed ones represent the field theory prediction in \eqref{eq:ham_two_attach}. The zoom focuses on the oscillating terms around $x=x_{S,2}$, showing that a more refined averaging procedure (see Eq. \eqref{eq:doubleAvgTrick}) suppresses them.}
    \label{fig:BiLocEnt}
\end{figure}

As for the bi-local terms, the entanglement Hamiltonian \eqref{eq:ham_two_attach} contains the bi-local operator  diagonal in the fermionic chiralities 
and also the one that mixes the fermionic chiralities, 
which are evaluated in different conjugate points obtained from $\tilde{x}_+$ and $\tilde{x}_-$ reported in \eqref{eq:xtildepm}.
To understand how these two bi-local terms appear from the continuum limit of $\hat{K}_A$, let us focus on the block $h^{(1,2)}$ as an example. Its continuum limit should produce bi-local terms in which the integration variable $x$ is in $[0, a]$ and the conjugate point is in $[b, c]$. Identifying which terms appear in the limit of $h^{(1,2)}$ can therefore be done by selecting the conjugate points that map the first interval into the second.
From the mappings in \eqref{eq:entTwoAttached} (see also Fig.~\ref{fig:entTwoAttached}), we see that $\tilde{x}_+([0, a])$ is a subset of $[b, c]$ and, as a consequence, the bi-local diagonal continuum limit \eqref{eq:limitBilocDiag} with $h^{(1,2)}$ is proportional to the bi-local diagonal operator $T^\text{bi-loc}_\text{diag}$ of \eqref{eq:qlocdiag} calculated in $\tilde{x}_+$.
On the other hand, from \eqref{eq:entTwoAttached} we see that $\tilde{x}_-([0, a])$ is a subset of the reflection of $[b, c]$ with respect to the boundary. Due to the presence of the boundary, these points are reflected (see Fig.~\ref{fig:entTwoAttached}) and therefore the bi-local mixed limit \eqref{eq:limitBilocMix} for the block $h^{(1,2)}$ is proportional to the bi-local mixed operator $T^\text{bi-loc}_\text{mix}$ of \eqref{eq:mixBilocOp} evaluated in $-\tilde{x}_-$.
Similar considerations can be applied to the other matrix blocks, and we finally identify that the continuum limits proportional to operators calculated in $\tilde{x}_+$ are
\begin{equation}\label{eq:xtildepiuEH}
\begin{split}
    &\mathcal{S}_+^\text{diag}(x) = \sum_{j} (-1)^{(j-i-1)/2} h_{i,j}^{(1,2)}, \;\;\; \quad x \in [0, a]\,,
\\
  &  \mathcal{S}_+^\text{mix}(x) = \sum_{j} (-1)^{(i+j-1)/2} h_{i,j}^{(2,2)}, \;\;\; \quad x \in [b, c]\,,
    \end{split}
\end{equation}
while those proportional to operators calculated in $\tilde{x}_-$
\begin{equation}\label{eq:xtildemenoEH}
\begin{split}
    &\mathcal{S}_-^\text{diag}(x) = \sum_{j} (-1)^{(j-i-1)/2} h_{i,j}^{(2,1)}, \;\;\; \quad x \in [b, c]\,,
\\
\rule{0pt}{1.2cm}
   & \mathcal{S}^\text{mix}_-(x) = \begin{cases}
        \sum_{j} (-1)^{(i+j-1)/2} h_{i,j}^{(1,2)},   \;\;\;\;\; &x \in [0, a]
        \\
        \rule{0pt}{.7cm}
        \sum_{j} (-1)^{(i+j-1)/2} h_{i,j}^{(2,1)},   &x \in [b, c]\,.
    \end{cases}
\end{split}
\end{equation}

 In Fig.~\ref{fig:BiLocEnt} we compare the two different limits \eqref{eq:xtildepiuEH} and \eqref{eq:xtildemenoEH} with the predicted bi-local weights $\beta^\text{loc }_\text{sym}(\tilde{x}_{\pm})/(x-\tilde{x}_{\pm})$ in \eqref{eq:ham_two_attach} for intervals of equal length $\ell_1 = \ell_2 = 64$ and $\ell_1 = \ell_2 = 192$, showing their collapse for different system sizes. In the top panel, we consider the bi-local weight 
of the operators calculated in $\tilde{x}_+$, where the continuum limit is given by \eqref{eq:xtildepiuEH} combined with the average in \eqref{eq:avgTrick}. Both the diagonal part and the mixed one converge to the field-theoretical result of \eqref{eq:ham_two_attach}: the former to the prefactor of $T^\text{bi-loc}_\text{diag}(x,\tilde{x}_+, 0)$, the latter to the prefactor of $T^\text{bi-loc}_\text{mix, vec}(x,-\tilde{x}_+, 0; \pi)$.

In the bottom panel of the same figure, we repeat a similar analysis for the bi-local weight of the operators calculated in  $\tilde{x}_-$, whose continuum limit is given by \eqref{eq:xtildemenoEH} 
with the average in \eqref{eq:avgTrick} adapted to this quantity.
We find also in this case a good agreement with the prefactor of $T^\text{bi-loc}_\text{diag}(x,\tilde{x}_-, 0)$ (blue/pink) and of  $T^\text{bi-loc}_\text{mix, vec}(x,-\tilde{x}_+, 0; \pi)$ (green/brown) in our prediction in \eqref{eq:ham_two_attach}.
Moreover, in the second interval, we observe a cross-over between the non-diagonal operator and the diagonal one, in correspondence of the point $x_{S,2}$ in \eqref{eq:selfpoint} (dotted grey line), confirming what we found in \eqref{eq:ham_two_attach}. We remind that $x_{S,2}$ is a self-conjugate point, i.e. one of the zeroes of the function $w_\text{sym}$, which for the subsystem $A$ under consideration is given by \eqref{eq:wsym}.
We can now explain the mechanism of this cross-over on the lattice. From the expression of $\mathcal{S}^\text{mix}_-(x)$ in \eqref{eq:xtildemenoEH}, when $x \in [b,c]$, 
the phase of the matrix element $h_{i,j}$ is a smooth function for $x < x_{S,2}$ while it is strongly oscillating for $x > x_{S,2}$. The opposite happens for $\mathcal{S}^\text{diag}_-(x)$. 
The goal of the averaging procedure of \eqref{eq:avgTrick} is precisely to eliminate the strongly oscillating contributions and therefore it makes possible to clearly see the cross-over. However, as we can see from Fig.~\ref{fig:BiLocEnt}, this averaging is not sufficient to completely remove the oscillations at the cross-over point, which moreover appear to be independent of the size of the subsystem. In the zoom of Fig.~\ref{fig:BiLocEnt}, we show that an additional averaging procedure which extends up to the next-to-nearest neighbours (i.e. not only nearest neighbours as in \eqref{eq:avgTrick}) is sufficient to eliminate these residual oscillations. 

It is given by 
\begin{equation}
\label{eq:doubleAvgTrick}
\overline{S}^{\,\mathrm{diag}}_-(i) \equiv  \frac{1}{6}\,\widetilde{\mathcal{S}}^{\mathrm{diag}}_-(i-2) + \frac{1}{6}\,\widetilde{\mathcal{S}}^{\mathrm{diag}}_-(i-1) + \frac{1}{3}\,\widetilde{\mathcal{S}}^{\mathrm{diag}}_-(i) + \frac{1}{6}\,\widetilde{\mathcal{S}}^{\mathrm{diag}}_-(i+1) + \frac{1}{6}\,\widetilde{\mathcal{S}}^{\mathrm{diag}}_-(i+2)\,,
\end{equation}
(where  $\widetilde{\mathcal{S}}^{\mathrm{diag}}_-(i)$ is defined by the combination \eqref{eq:avgTrick} 
with $\mathcal{S}^{\mathrm{diag}}_-(i)$ instead of $\mathcal{S}^{\mathrm{loc}}_-(i)$)
and by the same combination for $\overline{S}^{\,\mathrm{mix}}_-(i)$.
As anticipated, the zoom of Fig.~\ref{fig:BiLocEnt}
shows that using \eqref{eq:doubleAvgTrick}, i.e. red and orange symbols, the oscillations are suppressed  with respect to ones obtained through \eqref{eq:avgTrick}, i.e. green and blue symbols.

\subsubsection{Negativity Hamiltonian}
\label{sec-two-int-bdy-en}
\begin{figure}[t!]
\vspace{-.0cm}
    \centering
    \includegraphics[width=.7\textwidth]{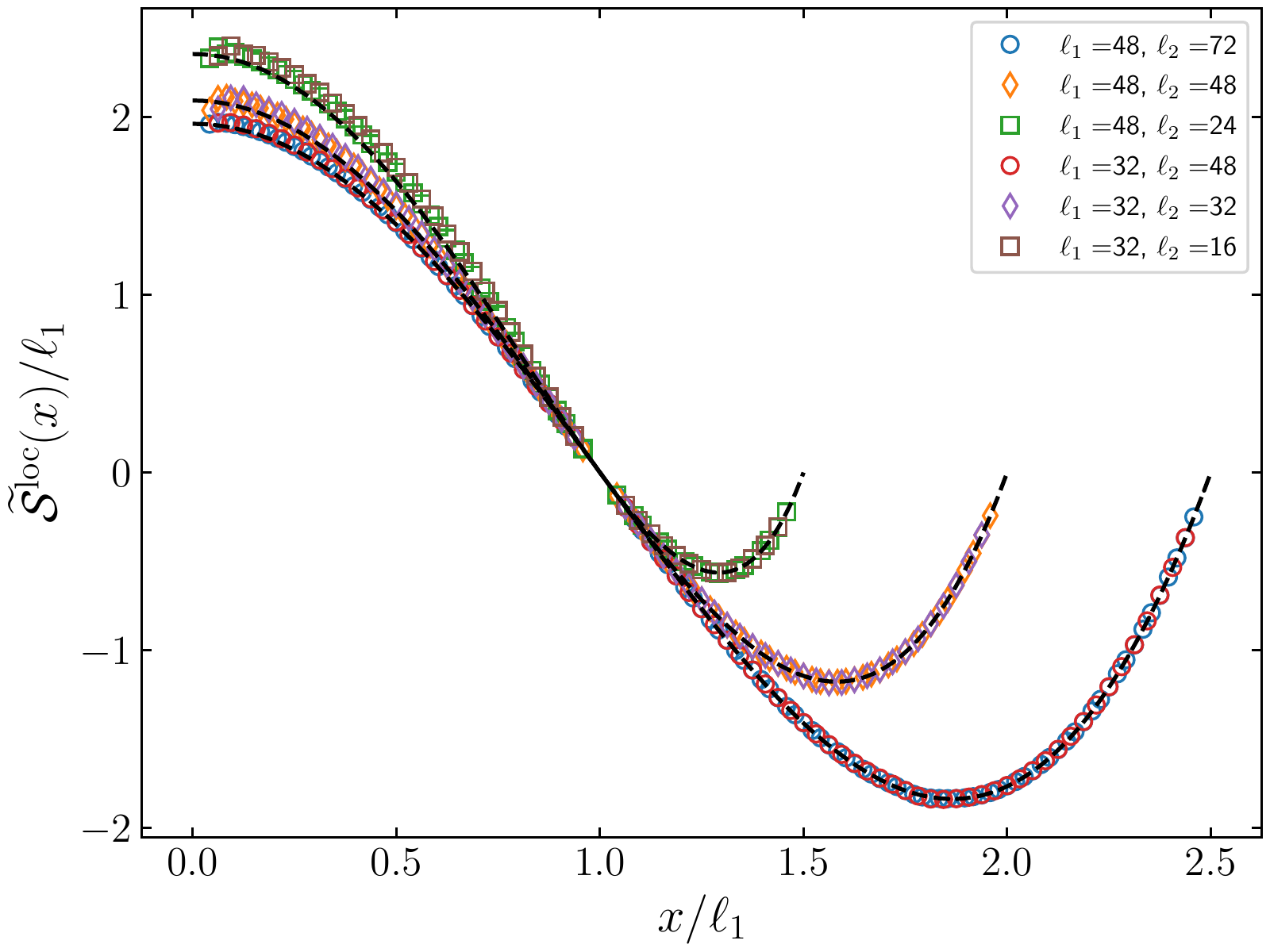}
    \vspace{-.2cm}
    \caption{Inverse effective temperature of the negativity Hamiltonian, rescaled with $\ell_1$ as a function of $x/\ell_1$. 
    The geometry we consider is $A=[1,\ell_1] \cup [\ell_1+1,\ell_1+\ell_2]$ for different values of the ratio $\ell_1/\ell_2 = 0.5, 1, 1.5$ and with a boundary at $x=0$. 
    The data points are obtained by applying \eqref{eq:avgTrick} to \eqref{eq:limitLocalRows} for the first interval and to \eqref{eq:limitLocalAntidiag} for the second one.
    The dashed curves correspond to the CFT expression \eqref{eq:betaR}.}
    \label{fig:LocNeg_diffRatios}
\end{figure}

\begin{figure}[t!]
\vspace{.8cm}
    \centering
    \includegraphics[width=1.0\textwidth]{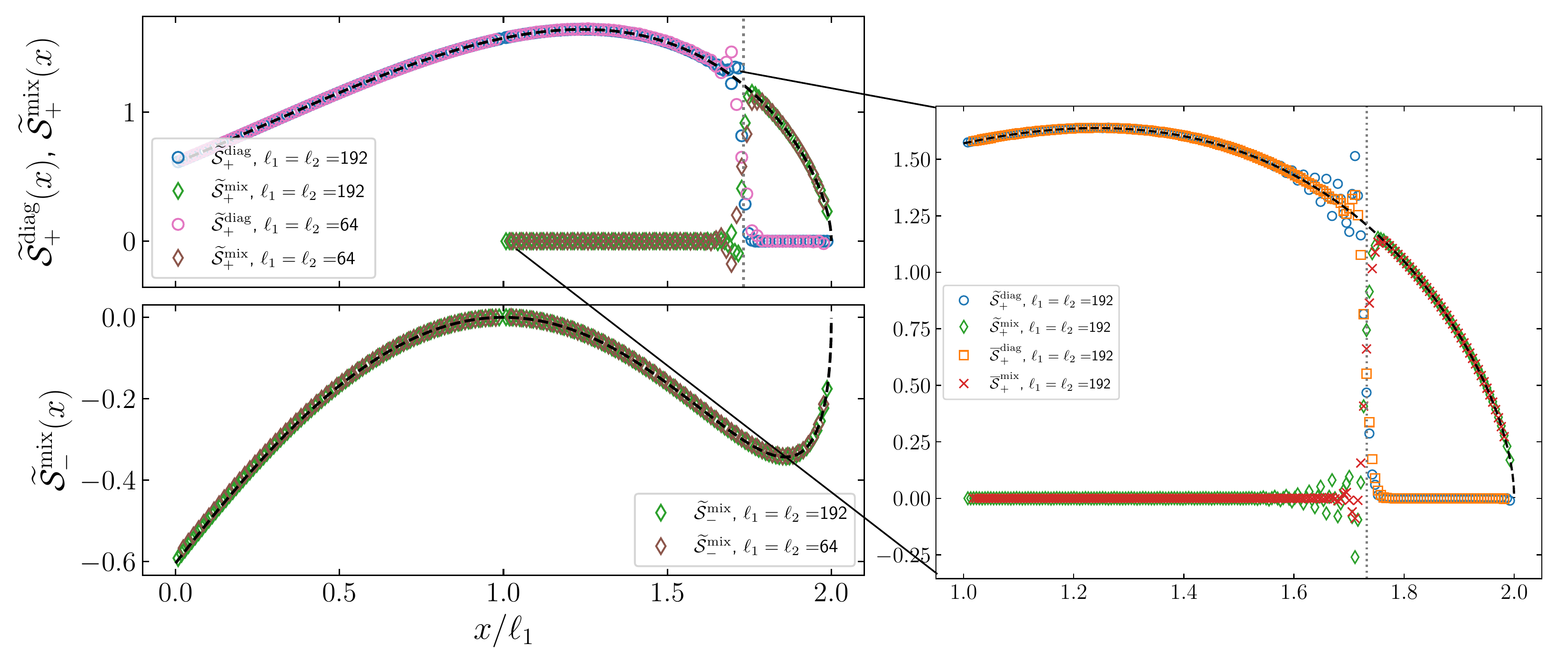}
    \vspace{-.6cm}
    \caption{
    Weight functions of the bi-local terms in the negativity Hamiltonian
    as a function of $x/\ell_1$ for the geometry $A=[1,\ell_1] \cup [\ell_1+1,2\ell_1]$ in the presence of a boundary at $x=0$
    (see \eqref{eq:nhTrip}) for $\ell_1=64$ and $\ell_1=192$.
    The bi-local weight functions for $\tilde{x}_+$ (top panel)  and $\tilde{x}_-$ (bottom panel) are shown.
    The green/brown and blue/pink colours distinguish between the non-diagonal and diagonal operators respectively, 
    by applying \eqref{eq:avgTrick} to \eqref{eq:xtildepiuNH} and \eqref{eq:xtildemenoNH}. 
    The vertical dotted grey line indicates the self-conjugate point $x^R_{S,2}$ (see the text below \eqref{eq:betaR}).
    The dashed  curves correspond to the field-theoretical prediction \eqref{eq:nhTrip} for the weight function of the bi-local terms. The zoom shows the suppression of the oscillations around $x=x^R_{S,2}$ using the more refined averaging procedure reported in \eqref{eq:doubleAvgTrick}.}
    \label{fig:BiLocNeg}
\end{figure}

As we discussed in Sec.~\ref{susec:NHinfinite}, also in the presence of the boundary the derivation of the continuum limit from the lattice negativity Hamiltonian is almost unmodified. Again, the only difference comes from the imaginary factors due to the partial transpose operation. For $A=[0,a] \cup [b,c]$, the appropriate factors can be read from the field-theoretical result in \eqref{eq:nhTrip}. By properly modifying \eqref{eq:xtildepiuEH} and \eqref{eq:xtildemenoEH} to take into account the transposition of the second interval, we find for the bi-local term calculated in $\tilde{x}_+^{R}$
\begin{equation}\label{eq:xtildepiuNH}
\begin{split}
    \mathcal{S}_+^\text{diag}(x) = \begin{cases}
        \; -\ii \sum_{j} (-1)^{(j-i-1)/2} \eta_{i,j}^{(1,2)}, \;\;\;   &x \in [0, a]
        \\
        \rule{0pt}{.7cm}
        \; \ii \sum_{j} (-1)^{(j-i-1)/2} \eta_{i,j}^{(2,1)},   &x \in [b, c]\,,
    \end{cases}
    \\
            \rule{0pt}{.9cm}
    \mathcal{S}^\text{mix}_+(x) = \, \ii \sum_{j} (-1)^{(i+j-1)/2} \eta_{i,j}^{(2,1)}, \;\;\;\; \qquad x \in [b, c]\,,
    \end{split}
\end{equation}
while for the one in $\tilde{x}_-^R$
\begin{equation}\label{eq:xtildemenoNH}
    \mathcal{S}_-^\text{mix}(x) = \begin{cases}
        -\ii \sum_{j} (-1)^{(i+j-1)/2} \eta_{i,j}^{(1,2)}, \;\;\;  &x \in [0, a]
        \\
                \rule{0pt}{.7cm}
        \sum_{j} (-1)^{(i+j-1)/2} \eta_{i,j}^{(2,2)},   &x \in [b, c]\,.
    \end{cases}
\end{equation}
We can now test our predictions for the negativity Hamiltonian in \eqref{eq:nhTrip}. Choosing $a=b=\ell_1, c=\ell_1+\ell_2$, in Fig.~\ref{fig:LocNeg_diffRatios}, we use the continuum limit in \eqref{eq:limitLocalRows} for the first interval adjacent to the boundary and \eqref{eq:limitLocalAntidiag} for the second one, together with the average in \eqref{eq:avgTrick}. For all the different length ratios $\ell_2/\ell_1 = 0.5, 1, 1.5$, we find good agreement with the analytical prediction for $\beta^{R \text{ loc}}_\text{sym}(x) $ in \eqref{eq:betaR}. The small discrepancy in the first interval can again be attributed to the effects of the sum along rows rather than along antidiagonals.

In Fig.~\ref{fig:BiLocNeg} we benchmark the bi-local term of the negativity Hamiltonian with $\ell_2=\ell_1$. In the top panel, we compare the continuum limit in \eqref{eq:xtildepiuNH} with $\beta^{R \text{loc}}_\text{sym}(\tilde{x}^R_+)/(x-\tilde{x}^R_+)$ in \eqref{eq:nhTrip}. The green/brown (blue/pink) symbols agree with the weight function that multiplies the operator $T^\text{bi-loc}_\text{mix, vec}(x, -\tilde{x}^R_+, 0; \pi)$ ($T^\text{bi-loc}_\text{diag}(x, \tilde{x}^R_+, 0)$). We also observe the cross-over between the diagonal and the non-diagonal operators in correspondence of the self-conjugate point $x_{S,2}^R = \sqrt{ca + bc - ab}$ (dotted grey line) (see discussion below \eqref{eq:betaR}). The zoom shows also here that the use of the additional average in \eqref{eq:doubleAvgTrick} suppresses the oscillations around $x_{S,2}^R$ with respect to \eqref{eq:avgTrick}. Finally, in the bottom panel we compare the continuum limit in \eqref{eq:xtildemenoNH} with $\beta^{R \text{ loc }}_\text{sym}(\tilde{x}^R_-)/(x-\tilde{x}^R_-)$ in \eqref{eq:nhTrip}, finding again good agreement.

\section{Conclusions}
\label{sec:conclusions}

We have studied the entanglement Hamiltonian of a multi-interval subsystem $A$ of a massless Dirac fermion on the half-line, 
generalising the result for the single interval found in \cite{mt-21}. 
The boundary condition can be implemented in two different ways, either preserving the charge (in the vector phase) or the helicity (axial phase).
Interestingly, while the entanglement entropy of this geometry is identical in the two phases, the entanglement Hamiltonian distinguishes between them. 
The entanglement Hamiltonians can be written as a sum of a local operator proportional to the stress-energy tensor and a bi-local one, which mixes the two chiral components of the Dirac field with a non-trivial dependence on the phase (see \eqref{eq:mixVecBilocOp} and \eqref{eq:mixAxBilocOp}). 
The latter operator is different for the two phases.
We have worked out explicitly the case $A=[0,a] \cup [b,c]$, i.e. two intervals with one adjacent to the boundary, from which we have recovered some well know limits as a consistency check.

The multipartite geometry offers also an ideal setting to compute the negativity Hamiltonian for free fermions recently introduced in \cite{mvdc-22}.
This is defined as the logarithm of the partially transposed reduced density matrix and represents an operatorial characterisation of entanglement in mixed states. 
After providing a construction scheme for a generic number of transposed intervals in the presence of a boundary, we focus on a tripartite geometry $[0,a]\cup [b,c]$, 
for which we report an explicit expression in \eqref{eq:nhTrip}.

We have also performed comparisons between our analytical predictions and the exact numerical computations in a free fermion chain by adapting the method discussed in  \cite{ETP19, EH-2, ETP22} for other cases.  
As for the local part of the entanglement and negativity Hamiltonian, the field theoretical results are obtained 
through a proper continuum limit, which includes also the long-range hopping terms (see Figs.~\ref{fig:LocEnt_diffInter} and \ref{fig:LocNeg_diffRatios}).
The weight function of the bi-local terms is recovered by a proper sum of the matrix elements multiplied by an oscillatory factor (see Figs.~\ref{fig:BiLocEnt} and \ref{fig:BiLocNeg}). 
In both cases, the numerical results show a perfect agreement with field theory.
 
There are a number of generalisations of the results presented here that are worth mentioning as outlooks. 
The first obvious one would be to investigate what happens on the lattice for a generic boundary condition as recently done for one interval in \cite{ETP22}.   
Another generalisation concerns the calculation of the entanglement and negativity Hamiltonians in the presence of slowly varying inhomogeneities, resulting, e.g. from 
external trapping potential or inhomogeneous initial states out of equilibrium. 
In this setting the CFT approach in curved space \cite{dsvc-17} can be used to describe universal quantities and it has been already employed for some entanglement Hamiltonians \cite{trs-eh-curved, fsc-22}.
An open problem is also the determination of entanglement and negativity Hamiltonians in the presence of a point-like defect which allows both reflection and transmission (a boundary condition is a purely reflective defect); the results for a single interval appeared already in \cite{mt2-21}. 
Finally, another natural question is what happens in higher dimensional boundary systems and how to recover the continuum limit from the lattice (as in \cite{jt-21} for the bulk case).

\section*{Acknowledgements}
We thank Filiberto Ares, Giuseppe Di Giulio, Viktor Eisler, Mihail Mintchev, Diego Pontello and Vittorio Vitale for useful discussions.
F.R., S.M. and P.C. acknowledge support from ERC under Consolidator grant number 771536 (NEMO).

\end{document}